\documentclass[twocolumn,showpacs,prd,groupedaddress,nofootinbib]{revtex4}

\usepackage{amsfonts,amsmath,amssymb,mathrsfs}
\usepackage{hyperref}
\usepackage{subfigure}
\usepackage{color}
\usepackage{graphicx}  
\usepackage{dcolumn}   
\usepackage{bm}        
\usepackage[english]{babel}

\def\a{\alpha}
\def\b{\beta}
\def\c{\gamma}
\def\r{\rho}
\def\s{\sigma}
\def\t{\tau}
\def\m{\mu}
\def\n{\nu}
\def\k{\kappa}
\def\th{\theta}
\def\g{\gamma}\def\G{\Gamma}
\def\L{\Lambda}\def\l{V}
\def\D{\Delta}
\def\la{\langle}
\def\ra{\rangle}
\def\o{\omega}\def\O{\Omega}
\def\d{\delta}
\def\p{\partial}

\def\oxthree{{\cal O}(x^3) }

\def\tV{\tilde{V}}
\def\tx{\tilde{x}}

\def\half{\textstyle{\frac{1}{2}}}

\def\bdoc{\begin{document}}
\def\edoc{\end{document}}
\def\bea{\begin{equation}}
\def\eea{\end{equation}}

\def\beq{\begin{eqnarray}}
\def\eeq{\end{eqnarray}}
\def\ben{\begin{enumerate}}
\def\een{\end{enumerate}}
\def\la{\langle}
\def\ra{\rangle}
\def\a{\alpha}
\def\g{\gamma}\def\G{\Gamma}
\def\d{\delta}\def\D{\Delta}
\def\e{\epsilon}
\def\z{\zeta}

\def\th{\theta}
\def\k{\kappa}
\def\l{\lambda}
\def\m{\mu}
\def\n{\nu}
\def\o{\omega}
\def\p{\pi}
\def\r{\rho}
\def\s{\sigma}
\def\t{\tau}
\def\L{{\cal L}}
\def\S{\Sigma }
\def\gsim{\; \raisebox{-.8ex}{$\stackrel{\textstyle >}{\sim}$}\;}
\def\lsim{\; \raisebox{-.8ex}{$\stackrel{\textstyle <}{\sim}$}\;}
\def\gtrsim{\gsim}
\def\lessim{\lsim}
\def\loc{{\rm local}}
\def\vm{v_{\rm max}}
\def\bh{\bar{h}}
\def\del{\partial}
\def\nab{\nabla}
\def\half{{\textstyle{\frac{1}{2}}}}
\def\fourth{{\textstyle{\frac{1}{4}}}}

\def\bD{{\bf D}}
\def\bE{{\bf E}}
\def\bF{{\bf F}}
\def\bB{{\bf B}}
\def\bP{{\bf P}}
\def\bV{{\bf v}}
\def\bv{{\bf v}}
\def\bx{{\bf x}}
\def\by{{\bf y}}
\def\bz{{\bf z}}
\def\ba{{\bf a}}
\def\bd{{\bf d}}
\def\bs{{\bf s}}
\def\bn{{\bf n}}
\def\bp{{\bf p}}

\def\O{\Omega}

\def\br{{\bf r}}
\def\bnab{{\bf \nab}}

\def\tE{\tilde{E}}
\def\tL{\tilde{L}}
\def\Horava{Ho\v{r}ava }

\def\oxtwo{\mathscr{O}\left(x^2\right)}
\def\oxthree{\mathscr{O}\left(x^3\right)}
\def\oxfour{\mathscr{O}\left(x^4\right)}
\def\oxfive{\mathscr{O}\left(x^5\right)}

\def\ph{\phantom}

\begin{document}

\title{Horizon entropy and higher curvature equations of state}
\author{Raf Guedens\footnote{rafguedens@gmail.com}, Ted Jacobson\footnote{jacobson@umd.edu}${}^{,a}$, Sudipta Sarkar\footnote{sudiptas@imsc.res.in}${}^{,a,b}$}
\affiliation{$^a$Center for Fundamental Physics,  University of Maryland, College Park, MD 20742-4111, USA}
\affiliation{$^b$Institute of Mathematical Sciences, Chennai, India}
\date{\today} 
\begin{abstract}
The Clausius relation between entropy change and heat flux has previously been used to 
derive Einstein's field equations as an equation of state. In that derivation the 
entropy is proportional to the area of a local causal horizon, and the heat 
is the energy flux across the horizon, 
defined relative to an approximate boost Killing vector.
We examine here whether a similar derivation can be given for 
extensions beyond Einstein gravity to include 
higher derivative and higher curvature terms. 
We review previous 
proposals 
which, in our opinion, are problematic or incomplete.
Refining one of these, we assume
that the horizon entropy depends on an 
approximate local Killing vector in a way that mimics the diffeomorphism Noether charge
that yields the entropy of a stationary black hole.
We show how this can be 
made to work if various restrictions are imposed on the nature of the horizon
slices and the approximate Killing vector. 
Also, an integrability condition on the assumed horizon entropy density must hold. 
This can yield field equations of a Lagrangian constructed algebraically from
the metric and Riemann tensor, but appears unlikely to allow for 
derivatives of curvature in the Lagrangian.
\end{abstract}  
\maketitle

\section{Introduction}

The notion of horizon entropy and thermodynamics
was first discovered for black holes in general relativity (GR) and 
quickly generalized to other sorts of horizons. 
The origin of this thermodynamic
behavior can be traced to local physics, and in a sense arises
from the nature of the vacuum. This led to the observation that
the Einstein equation can be derived as an equation of state
for local causal horizons 
in the neighborhood of a point 
`$p$' in spacetime, 
by imposing the Clausius relation between their entropy change
and the energy flux across them~\cite{Jacobson:1995ab}. 
In this paper we examine approaches to generalizing this
equation of state derivation to allow for higher derivative contributions 
to the entropy and field equations. 
The Einstein-Hilbert Lagrangian is only the lowest order
term (other than the cosmological constant) 
in a derivative expansion of generally covariant actions
for a metric theory, and the presence of higher derivative terms
is presumably inevitable.
Several approaches to including higher derivative terms
in the entropy and equation of state have been 
tried.  In our view none 
have fully succeeded, except in the case of ${\cal L}(R)$ theories.
Those theories are special however, since they 
are trivially related to general relativity 
coupled to a scalar field
by a field dependent 
conformal rescaling of the metric. Here we will explain the problems 
that arise with previous proposals, and 
propose a solution that adopts aspects of some of the
proposals. 

The solution differs in several ways
from the original derivation for GR, among which are: 
(i) the entropy is compared
on two horizon slices that share a common boundary, 
(ii) the bifurcation surface lies to the past of the terminal point $p$ and,  
(iii)  the entropy depends on the approximate Killing vector. In particular, it has the same dependence on the approximate Killing vector as have the Noether charges associated with a Lagrangian, in analogy with the Wald entropy~\cite{Wald:1993nt,Iyer:1994ys} 
for stationary black holes. Such dependence will be referred to as ``Noetheresque", but by itself does not make the entropy a Noether charge.  
It makes some sense that the entropy depends on the approximate Killing vector, because the latter determines the notion of stationarity and enters the definition of the heat flux. However, at present we can offer no statistical interpretation for this form of the entropy.  

The equation of state we can derive is consistent with local energy-momentum
conservation only if the leading order term in the entropy satisfies
an integrability condition. This condition is satisfied if the entropy arises from variation of a generally covariant function with respect to curvature. In other words, the entropy coincides with a Noether charge associated with a 
(particular type of) gravitational Lagrangian.
The need for such an integrability condition was anticipated in Ref.~\cite{Jacobson:1995ab},
since it was known that the entropy
of stationary black hole horizons has this 
form\cite{Visser:1993nu,Jacobson:1993vj,Iyer:1994ys}.
We have not been able to ascertain whether this is the only way to satisfy the integrability condition,
and in particular whether
field equations for Lagrangians involving derivatives of curvature can be obtained.

Besides the lack of a statistical interpretation,
and the dependence of the entropy on the local 
Killing vector, a strange feature of this approach is
that in the case of GR the entropy on a general 
horizon slice differs from the area at the same order 
as the relevant area changes in the Clausius relation.
So this approach seems to lose contact with some
of the original statistical 
motivation for the local Clausius relation.
On the other hand, it is quite analogous to the 
first law of black hole mechanics in generalized
gravity theories. Therefore it is not clear to us whether this
approach is purely formal, or maintains
some significance as true thermodynamics of 
spacetime.


\section{The Einstein equation of state}
\label{Einstein equation of state}

We begin with a review of the derivation of the Einstein equation
as an equation of state,
which emerges from black hole thermodynamics as follows.
General relativity and quantum field theory in a black hole
background imply the so-called ``first law of black hole 
thermodynamics"~\cite{Bekenstein:1973ur, Bardeen:1973gs, Hawking:1976de},
\beq\label{1stlaw}
d M -\O_Hd J=T_H d S_{BH},
\eeq
where $M$ is the black hole mass, $\O_H$ the 
angular velocity of the horizon, $J$ the
angular momentum, $T_H=\hbar\kappa/2\pi$ with 
surface gravity $\kappa$ is the Hawking temperature,
and $S_{BH}=A/(4\hbar G)$ with horizon area $A$ 
is the Bekenstein-Hawking entropy.
This relation can be viewed as a comparison between two
stationary black holes, but it also holds for small, slowly time-dependent
changes of a single black hole. Its validity in that setting hinges on
the fact that the evolution of horizon area is governed by 
spacetime curvature, which in turn is linked via Einstein's
equation to the energy flux. Specifically, it relies on the relation
$R_{ab}k^ak^b = 8\pi G T_{ab}k^a k^b$, where 
$R_{ab}$ is the Ricci tensor, $T_{ab}$ is the energy momentum
tensor of matter, and $k^a$ is a 4-vector tangent to the horizon
generating null geodesics.

The term ``first law" is actually a misnomer for (\ref{1stlaw}).
In thermodynamics that name refers to energy conservation,
$dU=\d Q + \d W$, where $dU$ is the internal energy change, 
$\d Q$ is the heat flow into and $\d W$ is the work done on 
the system. Instead, the thermodynamic nature of 
(\ref{1stlaw}) is the Clausius relation 
\beq
\d Q = T \d S 
\eeq
between the heat flow and the entropy change. In the black hole
context, energy that flows across the horizon is, in effect, 
``heat", since after crossing the horizon its microscopic nature 
is effaced for an outside observer. Since the horizon is
a causal barrier, it is a ``perfect dissipator"~\cite{Candelas:1977zz}. 
The heat flux is
\beq
\d Q = dM-\O_H dJ = \int_{\cal H}(-T_{ab}\chi^a) dH^b,
\eeq
where the integral is over the horizon and
$\chi^a=\del_t^a +\O_H\del_\phi^a$ is the 
horizon-generating Killing vector.\footnote{ We (unfortunately) 
use spacetime signature $({-}{+}\cdots{+})$.
The horizon integration measure is 
$dH_b=-k_b dV dA$, where 
$V$ is the affine parameter along the horizon generators, 
$k^b=(\partial_V)^b$ 
is tangent to the generators, and $dA$ is the area element of 
a constant $V$ horizon slice.}
That is, the heat is the
energy flux conjugate to the spacetime symmetry that translates along 
the horizon generators. The work term $\d W$ in the usual
first law of thermodynamics has no analog in the relation (\ref{1stlaw}).
Although ``first law" is not an appropriate name, we will
use it here since it is entirely standard terminology.

Close to a black hole horizon the Hawking temperature
becomes the Unruh temperature $\hbar a/2\pi$ for stationary, uniformly
accelerated observers, and the Clausius relation takes on
a local form (how local depends on how slowly 
the changes take place) whose validity requires that the 
Einstein equation hold, as mentioned above.
Moreover, there is good reason 
to believe this applies not just to black holes but to any causal 
horizon. (See e.g.\ \cite{Jacobson:2003wv} for a review of the arguments, 
and \cite{Amsel:2007mh} for further discussion and clarification.)

The idea of Ref.~\cite{Jacobson:1995ab} (see also \cite{Eling:2006aw} for a slight
reformulation) was that, conversely, the
Einstein equation can be derived, as an equation of state,
by requiring that the Clausius 
relation hold for the entropy of sufficiently small 
patches of all local causal horizons (LCHs) in spacetime.
Any such horizon $H$ is defined as the boundary of the 
past of a patch of $(D-2)$-dimensional spacelike surface,
where $D$ is the spacetime dimension. 
(LCHs will be defined precisely in what follows;
for now it suffices to remain somewhat vague.)
In that derivation 
the heat is taken as the boost energy flux, defined with respect to
an approximate local boost Killing vector field $\xi^a$ as
\bea \label{heatflux}
\delta Q = \int_{H}(-T_{ab}\xi^a) dH^b,
\eea
and the temperature is taken as the 
Unruh boost temperature $\hbar/2\pi$. This is natural since the
Minkowski vacuum state of quantum fields is thermal
with respect to the boost Hamiltonian at this temperature,
and any state looks like the Minkowski vacuum at sufficiently
short distances.
The entropy is taken to be $A/l_0^2$, where $l_0$ is some 
UV length scale. 
(A compelling case can be made that the origin of horizon
entropy is quantum entanglement of vacuum correlations
across the horizon. Since this is dominated by UV degrees 
of freedom, yet finite, dimensional analysis suggests that the 
leading order contribution should scale in this way.)
The change in horizon area in a small neighborhood of a
point $p$ can be related to the Ricci tensor via the
Raychaudhuri  equation. If the horizon is chosen so its
expansion and shear vanish at $p$, 
and $\xi^a$ is chosen so it too vanishes at $p$,
the Clausius relation is
then seen to require
$T_{ab}k^a k^b=(\hbar/(2\pi l_0^2))R_{ab}k^ak^b$
at every point in spacetime and for all null vectors $k^a$.
Together with energy conservation $\nab^a T_{ab}=0$ this 
implies the Einstein equation
\beq
R_{ab}-\half R g_{ab} -\Lambda g_{ab}=8\pi G_0 T_{ab},
\eeq
where the value of Newton's constant is determined by
the entropy density $1/l_0^2$ to be $G_0= l_0^2/4\hbar$, and $\Lambda$ is an undetermined cosmological constant.

\section{Previous approaches to including higher derivatives }

There have been a number of attempts to 
include higher derivative terms in
the equation of state derived from 
causal horizon entropy; 
all these attempts can be divided into two broad classes: The first deals with a specific theory, $L(R)$ gravity (see Refs.~\cite{Eling:2006aw, Elizalde:2008pv, Chirco:2010sw}), while
the second class studies a more general case(e.g. Refs.~\cite{Parikh:2009qs,Brustein:2009hy}).
The $L(R)$ gravity, where the Lagrangian depends only on the Ricci scalar, is equivalent to general relativity with an auxiliary scalar field \cite{Whitt:1984pd,Magnano:1987zz} and is therefore the simplest generalization possible. Even in that case, the derivation of the field equation from thermodynamics of local horizon meets considerable obstacles. 

Ref.~\cite{Eling:2006aw} considered the case when the horizon
entropy density is proportional to an arbitrary function 
$f(R)$ of the spacetime Ricci scalar $R$. 
As was done in the derivation of the Einstein equation, the approximate Killing vector $\xi$ is chosen to vanish at a terminal point $p$ on the horizon. The Clausius relation equates the entropy change to the heat flux $\d Q$, which is the 
flux of boost energy current $-T^{ab}\xi_b$ across the horizon. Since $\xi$ vanishes at the
terminal point, the rate of change of entropy with respect to the affine parameter must 
also vanish there. 
However, at a generic spacetime point $p$ the gradient $\nabla_a R$ is non-vanishing,
so the horizon entropy will have a nonvanishing rate of change
unless the change of $R$ is balanced by a change of horizon area.
Thus it is necessary to adjust the terminal surface of the 
horizon so that the expansion $\theta$ of the horizon
generators at the equilibrium point has the nonzero value
$\theta_p=-\dot{f}/f$. The Clausius relation involves a 
$\theta_p^2$ term
via the Raychaudhuri equation, and one ends up with a
field equation that is inconsistent with matter energy conservation
unless this term is identified as internal entropy 
production due to a bulk viscosity proportional to $f(R)$ and added to the entropy balance law. Hence, instead of equilibrium thermodynamics, a non-equilibrium approach is needed.  The resulting field 
equation is the one that follows from the Lagrangian 
$L$, where $f = dL/dR$. That is, $L$ is the 
Lagrangian for which $f(R)$ is the Wald entropy~\cite{Iyer:1994ys}. 
(The properties of Wald entropy will be briefly reviewed in section \ref{Wald entropy}.) 

In Ref.~\cite{Chirco:2010sw}, this entropy production term is interpreted as a separate contribution to the heat flux from the additional scalar degree of freedom present in $L(R)$ gravity. Then it is possible to derive the field equations using only reversible thermodynamics as in the case of general relativity. Although such an interpretation may offer a possible understanding of entropy production terms for the specific case of $L(R)$ gravity, it is unclear how to generalize that for a broader class of theories. 

In Ref.~\cite{Elizalde:2008pv} it was proposed that the need for internal entropy production 
could be eliminated by adopting the instantaneous boost invariant
(IBI) prescription for dynamical horizon entropy proposed in \cite{Iyer:1994ys}.
This prescription goes as follows. In the neighborhood of a spatial slice $\Sigma$
of a causal horizon one constructs, by a unique recipe of dropping selected
terms in a Taylor expansion of the true metric in an adapted coordinate
system, a new spacetime metric that has an exact boost Killing field for which 
$\Sigma$ is a fixed point set. 
The entropy density of the slice $\Sigma$ was taken in \cite{Elizalde:2008pv} 
to be a scalar formed from the IBI metric associated to 
$\Sigma$. In particular the case with entropy density $f(R)$ was 
discussed, but it is not clear to us that all contributions to 
the change in this entropy were taken into account. The difficulty arises
because the IBI metric changes with the horizon slice, so the 
$R$ on each slice is defined with respect to a different metric.
Also, even if the field equation arrived at in \cite{Elizalde:2008pv} 
were correct, it refers to the curvature of the IBI metric, and it remains unclear 
how this is related to the curvature of the original metric. \\

Next we would like to discuss attempts to obtain the field equation from local horizon thermodynamics beyond $L(R)$ gravity.
Ref.~\cite{Brustein:2009hy} starts from 
the Wald entropy formula~\cite{Iyer:1994ys} for a stationary 
black hole, but applied to a local horizon. 
A formula for the entropy of a [$(D-1)$-dimensional] patch of
a local horizon is written as a surface integral over the boundary
of the patch. It is unclear to us why an entropy should be assigned
to a ($D-1$)-patch rather than only to a ($D-2$)-slice of the horizon. Moreover, 
in the case of a stationary black hole horizon this
integral vanishes, whereas the entropy of the horizon surely does not
vanish. Hence this quantity cannot be interpreted as the entropy
of the horizon. The variation of this entropy is then considered, and a
formula is written in terms of an integral of a covariant directional
derivative of part of the integrand of the previously mentioned integral.
It is unclear to us why this expression should be interpreted as the
change of entropy. A similar formula is written for
the boost energy flux across the horizon. 
Also, the binormal on horizon cross sections is identified with the covariant derivative of the approximate Killing vector. Whereas this can certainly be satisfied to lowest order, 
given the small changes involved in the Clausius relation it
would be necessary to check whether
neglecting the differences at subleading order is justified. Another issue is that
the approximate Killing vector $\xi^a$ in these calculations is treated as if 
it satisfies the Killing identity $\nabla_a \nabla_b \xi_c = R^d_{\ph{d}abc} \xi_d$ exactly. 
 Although this is not possible in general, we will show in the present paper that the identity can be satisfied to the required order, provided one restricts to a narrow neighborhood of a particular horizon 
generator.

In Ref.~\cite{Parikh:2009qs} the proposal is to define the entropy of
horizon slices as the integral of a Noether potential formally identical to a potential introduced in~\cite{Iyer:1994ys}, except that in the former case no matter fields are included in the Lagrangian with which the potential is associated. 
Using Stokes' theorem,
the entropy change between two slices of the horizon 
is then expressed as an integral of the correspondig Noether current
over the enclosed horizon patch. However, since slices of a LCH have boundaries, the entropy change also involves a surface integral over the null outer boundary,  
unless the two slices have their boundary in common. This 
contribution was missed in Ref.~\cite{Parikh:2009qs}. In addition, as in 
Ref.~\cite{Brustein:2009hy}, the Killing identity was used without justification 
at an order it cannot be expected to hold in a generic spacetime.   
Thus this derivation too is not complete.  The present paper starts from
a similar but more general form for the entropy, and 
fills these gaps in the argument by 
studying slices that do have their boundary in common and thus enclose a compact patch of the horizon.
\\

Finally we note that Refs.~\cite{Padmanabhan:2009ry, 
Padmanabhan:2009jb} (see also
\cite{Padmanabhan:2009vy} and references therein) discuss a  
thermodynamic interpretation of gravitational field equations beyond  
general relativity in terms of a local entropy balance law. 
A matter entropy flux across the horizon is
associated with a small spatial volume and is related to
the boost energy by the Clausius relation. This is then
equated to a gravitational entropy change in the volume,
which is constructed from the Noether current associated
with a Lagrangian $L[g_{ab},R^{a}_{\ph{a}bcd}]$. This balance law at a point is then shown
to imply that the gravitational field equations hold, thus
giving a thermodynamical interpretation of those equations.
The approach we will pursue in this paper
uses similar ingredients, but we have different objectives. 
We start by assigning a horizon entropy functional to a 
small but finite sized horizon slice,
and investigate what properties it must have if it is to satisfy 
the Clausius relation with the matter boost energy flux.

\section{Impossibility of generalizing the non-equilibrium approach} \label{Impossibility}

A natural question to ask is whether the non-equilibrium approach of Ref.~\cite{Eling:2006aw} can be extended beyond theories for which the horizon entropy density $s$ is a function only of the Ricci scalar. 
In this section it will be shown that the non-equilibrium approach of Ref.~\cite{Eling:2006aw} cannot be extended beyond the results found there.
For this discussion, we adopt the geometric set up of \cite{Jacobson:1995ab,Eling:2006aw} and we assume
that the entropy associated with the LCH takes the form
\beq
S = \int s~ dA,
\eeq
where $s$ is some arbitrary scalar function
and the integration is over a $(D-2)$-dimensional spacelike slice of the horizon
with `area' element $dA$.
The change in entropy from one horizon slice to another is
\beq
\delta S =  \int \left ( \frac{d s}{d \lambda} + \theta s \right) d\lambda ~dA,
\eeq
where $\lambda$ is an affine parameter on the local horizon generators
and $\theta=d(\ln dA)/d\l$ is the expansion of the generators.

The Clausius relation asserts that the entropy change
$\d S$ is equal to $\delta Q/T$,
where 
$T=\hbar/2\pi$ is the boost temperature, $\d Q$ 
is the heat flux (\ref{heatflux}) through the horizon and
$\xi^a$ is the approximate Killing vector which vanishes at the final equilibrium
point $p$. Since the heat integrand vanishes at $p$, the Clausius relation
can 
hold only if the $\d S$ integrand also vanishes there. This requires that
the expansion at $p$ is nonzero, to wit
\beq
\theta_p = - \frac{1}{s}  \left. \frac{d s}{d \lambda} \right \vert_p .
\eeq
In the case of GR $s$ is a constant, so this condition states that the horizon
must have vanishing expansion at $p$.
The Raychaudhuri equation can be used to find $\theta(\l)$.
If the horizon shear does not vanish at $p$ we must add internal
entropy production due to shear viscosity to the
Clausius relation \cite{Eling:2006aw}.
Assuming the shear at $p$ vanishes, we have
$\theta=\theta_p +\lambda(-\theta_p^2/(D-2) -R_{ab}k^ak^b)+O(\l^2)$,
where $k^a$ is the affine horizon tangent vector
$(\mathrm{d}/\mathrm{d}\l)^a$, and $\l_p=0$.

Now we require that
the Clausius relation hold for all local horizons through $p$,
i.e.\ for all $k^a$. If the entropy density is independent of $k^a$,
as would be the case if it is a spacetime scalar, then we obtain
the tensorial equation
\beq
s R_{ab} - \nabla_a \nabla_b s + \frac{D-1}{D-2} s^{-1} s,_{a} s,_{b}+ \Psi g_{ab} = \frac{2 \pi}{\eta} T_{ab},
\label{generaleq}
\eeq
where $\Psi$ is some scalar function.
This function can be determined by imposing energy conservation,
i.e. $\nabla^a T_{ab}=0$, which yields the condition
\beq\label{dPsi}
\Psi,_{a} = - \frac{1}{2} s R,_{a}  +\partial_a \Box s - \left (\frac{D-1}{D-2} s^{-1} s,_{a} s,_{b}\right)^{;b}.
\eeq
Since $\Psi_{,a}$ is the gradient of a scalar, a solution exists only
if the right hand side is also the gradient of a scalar.

In the case of GR $s$ is a constant, so $\Psi=-sR/2$ is the unique solution
up to a constant (the cosmological constant).
If $s=s(R)$ is a function only of the Ricci scalar, then the first term on the
right hand side of (\ref{dPsi}) is a gradient, but the last term is not.
In the non-equilibrium approach of Ref.~\cite{Eling:2006aw}
the Clausius relation is replaced by
an entropy balance relation $\d S =\d Q/T+\d S_i$, where
$\d S_i$ is an internal entropy production term due to bulk viscosity,
which is proportional to the square of the expansion and which cancels the
last term in (\ref{dPsi}).
This yields an equation of state that coincides with the field equations
for $L(R)$ gravity.
For more general spacetime
scalar entropy densities, e.g.\ $s(R_{abcd})$, even the first term is not
a gradient, so there is apparently no way to satisfy an entropy balance
law, even allowing for internal entropy production.
Hence, the methods devolved in \cite{Jacobson:1995ab}
and  \cite{Eling:2006aw} can not be generalized beyond $L(R)$.

The situation is even more problematic if the entropy density is not
a spacetime scalar but, for example, is constructed from
the intrinsic curvature of the $(D-2)$-dimensional horizon slice, as in
Lovelock gravities \cite{Jacobson:1993xs}.
In that case, we do not even have a tensorial relationship like Eq.(\ref{generaleq}). The
Raychaudhuri equation then does not seem to be of any help at all,
so a very different approach is needed. In this paper we will present such an alternative approach, which involves an entropy density that depends explicitly on the
Killing vector, a different choice of horizon slices, and
the application of Stokes' theorem.


\section{New choice of approximate Killing vector}
\label{New choice}

\begin{figure}
  \begin{minipage}[b]{4cm}
    \includegraphics[width=4cm]{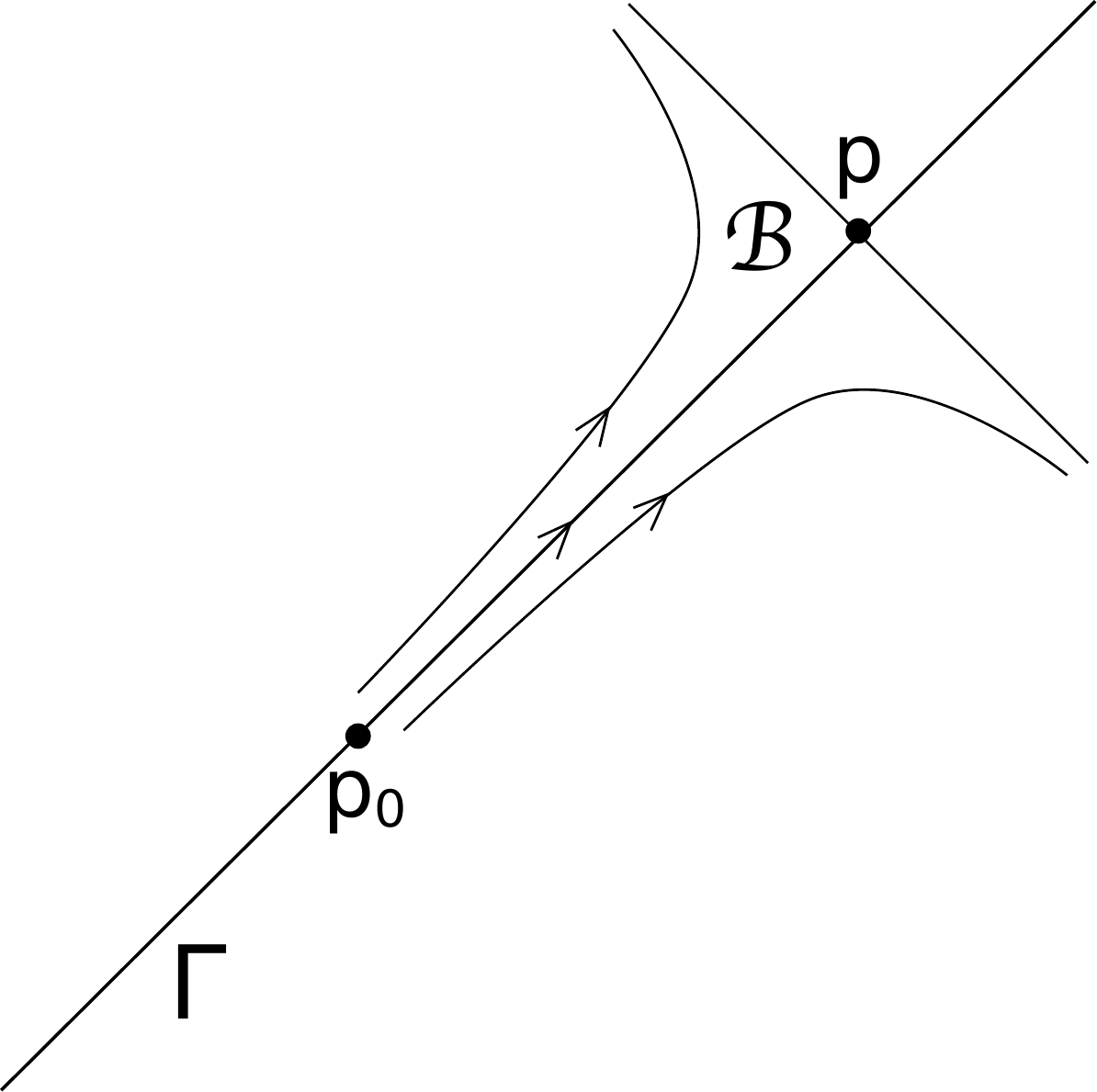}  
  \end{minipage}
  \begin{minipage}[b]{4cm}
    \includegraphics[width=3.3cm]{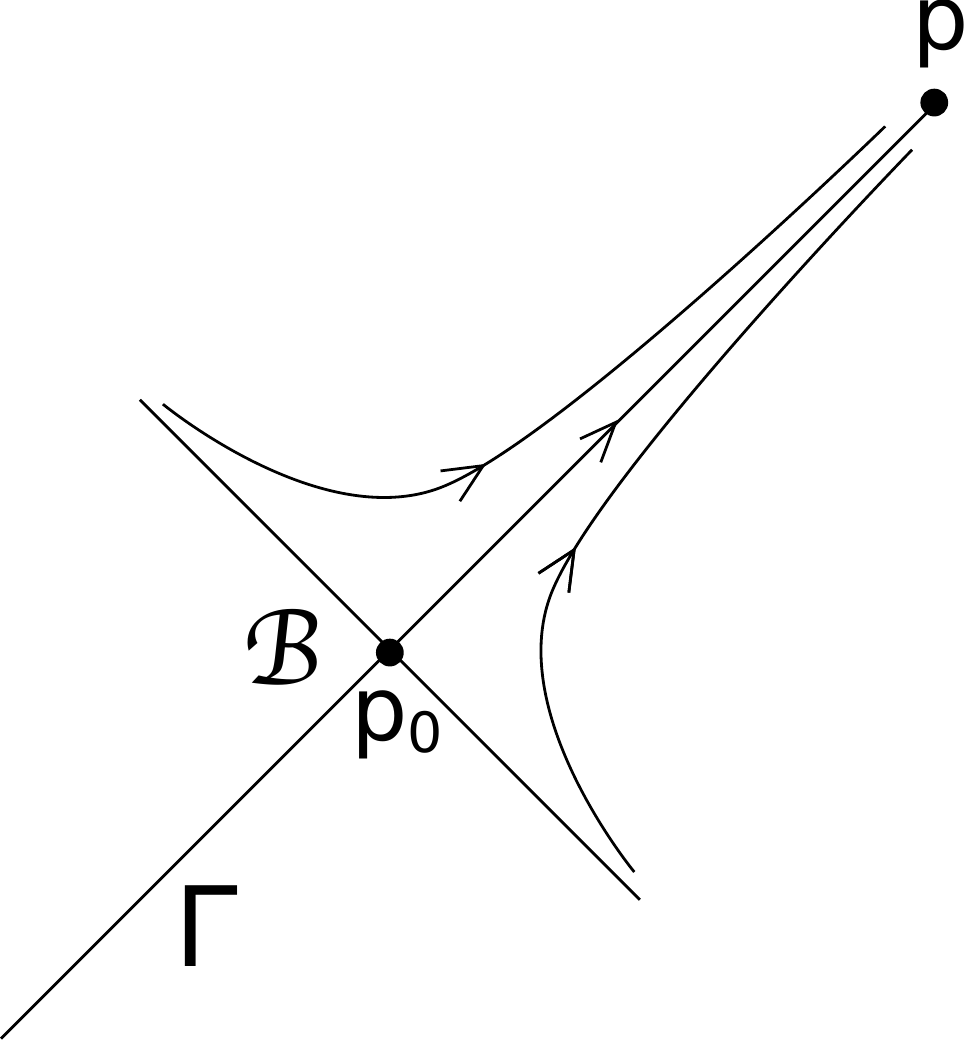}  
  \end{minipage}
  \caption{\label{setups}
  Causal spacetime diagram of the local causal horizon (LCH) and Killing field in the old (left) and new (right) setups. Each point in the diagram represents a patch of spacelike 2-surface. The boundary of the past of a patch of 2-surface through $p$ defines the LCH. The arrows indicate the flow lines of the local Killing field $\xi^a$, which is of third order on the bifurcation surface $\mathcal{B}$, 
and which vanishes at $p$ (left) or $p_0$ (right). The central horizon generator $\G$ runs from $p_0$ to $p$. The ``heat" is the boost energy flux across the horizon. 
In the new setup the Killing vector is timelike within the horizon and spacelike beyond,
while the old setup is the opposite.}
\end{figure}

It turns out that the new approach to the thermodynamic
derivation of field equations, even in the case of general relativity, 
calls for a small but essential adjustment in the choice of approximate 
Killing vector.
This adjustment amounts to locating the bifurcation surface\footnote{In the current context, the term ``bifurcation surface" is used somewhat loosely, mainly to fix attention to the analogous role played by the true bifurcation surface of a quasi-stationary black hole in the physical process version of the first law~\cite{Wald:1995yp}. What we mean is a slice on which the components of the approximate Kiling vector are third order in an appropriate coordinate system (see Sec.~\ref{LKV}). Accordingly, the approximate Killing vector vanishes at the center of the bifurcation surface (see Fig. \ref{setups}).}
at a time earlier than the terminal point, 
rather than being coincident with the terminal point as originally formulated in Ref.~\cite{Jacobson:1995ab}. 
The change also makes the application of the Clausius relation 
more closely analogous to the first law of black hole mechanics~\cite{Wald:1995yp} 
and resolves an 
uncomfortable aspect of the earlier derivation.
In this section we explain the thermodynamic motivation for 
the new choice of approximate Killing vector, and show how the
original derivation of Ref.~\cite{Jacobson:1995ab} is modified by this choice.
The role in the new approach will be explained in Sec. \ref{Clausius LCH}.

The role of the approximate Killing field $\xi^a$ 
in the approach of Ref.~\cite{Jacobson:1995ab}
is to define the heat in the Clausius relation (\ref{heatflux}). The bifurcation surface of this
Killing field was previously taken to coincide with the future boundary
through $p$,  
whereas in the first law of global horizon
mechanics the bifurcation surface lies to the {\it past} of the
perturbation.
Moreover, the approximate Killing field was spacelike
rather than timelike in the region outside the horizon, so it was
necessary to think of the heat as going into the reservoir 
{\it behind} the horizon (see Fig.~\ref{setups}).\footnote{We use the
term ``outside" to refer to the region accessible to observers 
to the past of $p$, like 
the observers outside a black hole horizon. For a cosmological
horizon, the standard terminology would be opposite: outside 
a cosmological horizon 
is the region that can {\it not} be seen.}
However, this reservoir is not observable
on the outside, so should play no role in the outside thermodynamics.
It seems much more satisfactory to place the bifurcation surface to the
past, so that the Killing field is timelike outside the horizon, and 
the reservoir can be thought of in direct analogy with the 
thermal atmosphere of a black hole, below the stretched horizon.\footnote{This same point was made recently
in Ref.~\cite{Yokokura:2011za}, with reference to the choice of the ``observer" who is defining the heat flux.}

It appears at first that modifying the location of the 
bifurcation point of the Killing vector will change the 
heat flux and, in the original approach of Ref.~\cite{Jacobson:1995ab}, ruin the derivation of the Einstein equation 
from the Clausius relation. However, that is not what happens.
The old Killing vector on the generator through $p$ 
(see Fig. \ref{setups})
was $\xi^{\rm old}_b=-\lambda k_b$, where 
$\lambda$ is the affine parameter that vanishes at 
$p$. The new Killing vector vanishes instead at $p_0$ and is given by 
$\xi_b=(\lambda-\lambda_0)k_b$, 
where $\lambda_0$ is the value of $\lambda$ 
at $p_0$.
The corresponding boost energy 
currents are thus
\begin{eqnarray}
-T^{ab}\xi^{\rm old}_b&=&\lambda T^{ab}k_b\\
-T^{ab}\xi_b&=&(\lambda_0-\lambda)T^{ab}k_b
\end{eqnarray}
Although these differ, their integrals from 
the bifurcation point to $p$ are the same, since
\begin{equation} \label{generator_integral}
\int_{\lambda_0}^0 \lambda\, d\lambda =-\lambda_0^2/2= \int_{\lambda_0}^0 (\lambda_0-\lambda)\, d\lambda.
\end{equation}
With the heat defined using $\xi^a$, applying the Clausius relation to the horizon interval $\left[\lambda_0,0 \right]$ 
in the limit $\lambda_0 \rightarrow 0$ thus yields the field equations as described above. This corresponds
to a transition between a stationary state at $\lambda=\lambda_0$ (where 
$\xi^a$ vanishes) and one at $\lambda=0$
(where the expansion and shear vanish). 
In analogy to the physical process version of the first law of black hole mechanics, the Clausius relation holds only when applied to this entire interval.

Note that in the approach of Ref.~\cite{Jacobson:1995ab}, with the new choice of Killing vector, one must still impose the condition that the expansion $\theta$ vanishes at $p$, because that condition is used in applying the Raychaudhuri equation to obtain the area change. By contrast, in the Noetheresque approach of the present paper the expansion at $p$ will turn out to play no role whatsoever.


\section{Properties of the local horizon and Killing vector} \label{geometry}

In this section we spell out the detailed geometric construction and properties 
of the local causal horizon (LCH) and Killing vector.

\subsection{Local causal horizon}
We define a local causal horizon $H$ as follows.
Consider any spacetime point $p$ in a $D$-dimensional spacetime,
and let $\Sigma_p$ 
be any small patch of spacelike $(D-2)$-surface
through $p$.
The boundary of the past of $\Sigma_p$ in the neighborhood of 
$p$ has two components, each of which is a null surface generated by 
a congruence of null geodesics orthogonal to $\Sigma_p$.
The LCH $H$ is defined as one of these components. 

As mentioned in the previous section, it will not matter for the present approach whether the congruence is stationary at $p$. However, if $p$ is to be a stationary point, the expansion and shear of
this congruence must vanish there. That is, the extrinsic curvature of $\Sigma_p$ must vanish at $p$, 
which is equivalent to saying that $\Sigma_p$ is generated by geodesics at $p$. 
For concreteness we will go further and assume that $\Sigma_p$ is fully generated by geodesics emanating
from $p$.
This will be convenient for studying concrete examples of the approximate Killing vector and of entropy values.

\subsection{Null normal coordinates}\label{sec:NNC}

To establish the existence of an approximate Killing vector
with the required properties,
we will employ
a ``null normal coordinate" (NNC) system~\cite{RafNNC}
adapted to the LCH. This is an explicit realisation of the coordinate systems first introduced in~\cite{Kay:1988mu, Iyer:1994ys} and is defined as follows. At a spacetime point $p$ 
an orthonormal set of $(D-2)$ spacelike vectors $\{e^a_A\}$, $A = 1,2,..,(D-2)$, 
is chosen, and  a $(D-2)$-surface $\Sigma_p$ is generated by geodesics 
with tangent vectors at $p$ in the space spanned by $\{e^a_A\}$. The point reached on such a
geodesic at unit affine parameter is assigned the coordinates $x^A$, when $x^A e^a_A$ is the tangent vector in that parametrisation at $p$.
This defines standard Riemann normal coordinates
on $\Sigma_p$  based at $p$. 
A pair of future pointing 
null vector fields $(k^a,l^a)$ normal to $\Sigma_p$ is chosen 
on $\Sigma_p$,
normalized such that $k_a l^a=-1$.
Each point $r$
in a small enough spacetime 
neighborhood of $p$ lies on a unique 
geodesic orthogonal to $\Sigma_p$ at some point $q$.
Let the tangent to that geodesic at $q$ be given by 
$V k^a + U l^a$ when $r$ lies at unit affine parameter from $q$. 
The NNCs of $r$ are then defined by\footnote{We use greek indices, and upper case latin indices, for coordinates and coordinate components in the NNC coordinate system. Lower case latin indices are reserved for abstract indices on tensors.}
\beq
x_r^\a= (U,V, x_q^A).
\eeq           
These coordinates are defined uniquely up to a rotation of the
$(D-2)$-frame at $p$ and a $q$-dependent rescaling of $k^a$, 
with inverse rescaling of $l^a$. 
The horizon $H$ is the surface $U = 0$, restricted to $V \leq 0$. 
In particular, the central horizon generator $\Gamma$ is the coordinate curve $U=0$, $x^A=0$, $V \leq 0$, with bifurcation point
$p_0 \in \Gamma$ at $V=V_0 < 0$. 
Note that on $H$ the coordinate $V$ is an affine parameter along the null generators.  
The details of the construction, including the derivation of  
metric coefficients and Christoffel
symbols, are presented in Ref.~\cite{RafNNC}. It is also shown there that
the ambiguity in the choice of $k^a$ can be exploited to 
 make the coordinates locally inertial at $p$ and to further
specialize the properties of the metric components. A preferred choice is made by starting with the $D$-dimensional Riemann normal
coordinates at $p$ and adjusting them. This also induces a particular choice of affine parameter $V$.

\subsection{Horizon slices}\label{sec:slices}

The Clausius relation applied to a horizon refers to the entropy change 
$\d S$ between two times. Those times are spacelike hypersurfaces, one to the future of the other, which intersect the horizon in two slices. For a black hole with a compact horizon, the two 
slices may bound a cylindrical region of the horizon.

\begin{figure}
\centering
\includegraphics[width=8cm]{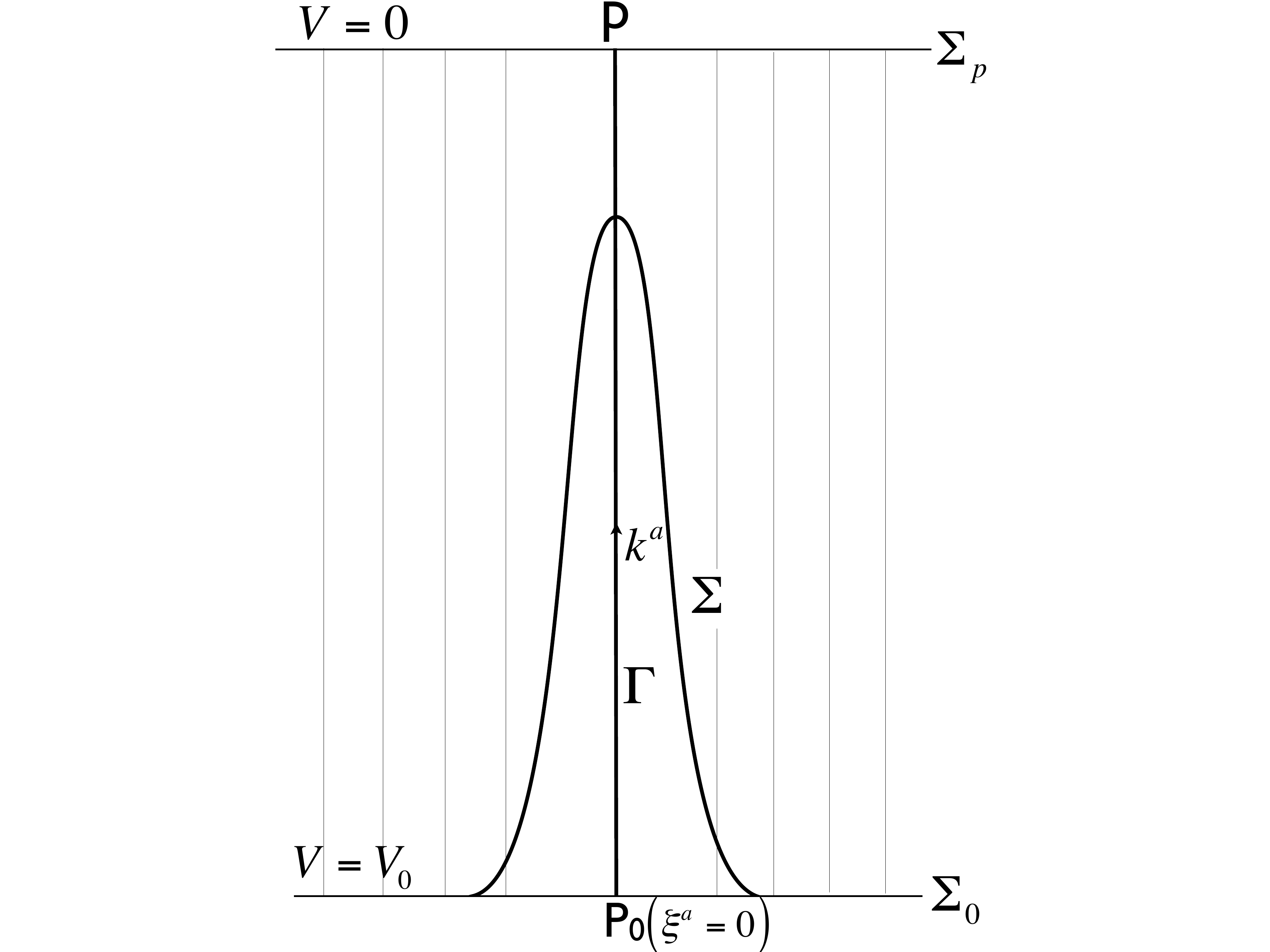}
\caption{\small The local causal horizon is part of 
the boundary of the past of $\Sigma_p$, a
spacelike $D-2$ dimensional surface.
The vertical lines are horizon generators with affine parameter
$V$ and tangent vector $k^a$, 
and $\Gamma$ is the central generator. 
$\Sigma_p$ is geodesic at $p$, so the 
expansion $\theta$ and shear vanish at $p$.
The horizontal lines represent constant $V$ slices.
The local Killing vector vanishes at the bifurcation point
$p_0$. 
The entropy
is compared on two slices, $\Sigma_0$ and $\Sigma$,
which differ only in a compact region.}\label{fig:slices}
\end{figure}

For a local causal horizon, the considered process
must be local, since the LCH is not even well defined 
except in a small neighborhood of the terminal point
$p$. To localize the process we can restrict attention
to cases where, rather than pushing the spacelike hypersurface forward in time everywhere, 
it is deformed to the future only in a small 
neighborhood of the bifurcation point 
$p_0$.\footnote{Hypersurface deformations of 
this sort were considered in Ref.~\cite{Wall:2011} in the context of 
proving a local form of the Generalized Second Law.} 
Then the two corresponding horizon slices also coincide
everywhere except in a small region. Parts of 
two such slices, $\Sigma_0$ and $\Sigma$, are 
depicted in Fig.~\ref{fig:slices}. $\Sigma_0$
corresponds to the $V=V_0$ surface
and $\Sigma$ lies to the future. If we
truncate the horizon slices outside the
region where they differ, their union 
$\Sigma_0 \cup \Sigma$ forms the 
closed boundary of  a patch of the horizon.
This will allow the difference of the entropies
on the two slices to be computed using Stokes'
theorem. 

In fact, we shall need to further restrict
the choice of horizon cuts so that they
bound a {\it narrow} region of the horizon,
because only then can the Killing identity 
be satisfied to sufficient accuracy for
the approximations made in
the derivation of the equation of state
to be valid\footnote{An alternative option would be to restrict to a horizon patch that is symmetric in the transverse coordinates, to cancel contributions from terms linear in those coordinates. However, we consider this symmetry restriction to be artificial.}. By ``narrow", 
we mean that the ratio of the width in $x^A$
to the length in $V$ goes to zero
in the limit as $p_0$ approaches $p$.

\subsection{Local Killing vector} 
\label{LKV}
Next we define precisely the approximate Killing vector $\xi^a$ that plays
a central role in defining both the heat flux and, in the Noetheresque 
approach of Sec. \ref{Noetheresque}, the entropy density.
We will refer to this vector field as the ``local Killing vector".

Of course a general curved spacetime has no Killing vectors. Nevertheless,
in a small enough neighborhood of any point, any spacetime is approximately
flat. In particular, in local inertial 
coordinates at $p$, the metric components
take the form $g_{\a\b}=\eta_{\a\b} + O(x^2/L^2)$, where $\eta_{\a\b}$ is the Minkowski
metric, $x$ denotes the coordinates, 
and $L$ characterizes the shortest radius of curvature of the spacetime at $p$.
(Here and below we use greek indices to refer to components in a particular
coordinate system, reserving lower case 
latin indices for abstract tensor indices.)
An approximate boost generator $\xi^a$ can be defined in terms of local inertial coordinates by the formula for 
an exact flat spacetime boost generator, e.g.\ $(x,t,0,0)$ in Minkowski coordinates.
This vector satisfies the Killing equation 
\beq\label{Keq}
\nabla_a \xi_b + \nabla_b  \xi_a = 0
\eeq
exactly 
at $p$, 
and to $O(x)$ near $p$, 
but in general $\nabla_{(a} \xi_{b)}$ will have $O(x^2)$ terms. 
This is adequate for our purposes.

However, our application of the Clausius relation is also sensitive to the $O(x)$ part of 
the Killing identity 
\beq \label{Kid}
\nabla_a \nabla_b \xi_c = R^d{}_{abc} \xi_d.
\eeq
For a true Killing field, this identity follows from 
the Killing 
equation (\ref{Keq}). 
Conversely, the Killing identity implies the Killing equation if the latter 
holds at one point, because of the antisymmetry of the Riemann tensor
in the last index pair.
Our computations will rely on this identity being
satisfied at $O(x)$, but
for the approximate Killing field defined above 
it will generally not hold at that order.
We can try to modify the definition of this vector field
so as to satisfy both the Killing equation and the 
Killing identity at $O(x)$,
but in fact this is not possible. 
However,
narrowing our sights, 
$\xi^b$ can be chosen
so that the Killing identity 
holds to this order (in fact it can be chosen to hold exactly) 
on the single horizon generator $\Gamma$
that ends on 
the terminal point $p$, and 
this turns out to be good enough.
In effect, the local Killing symmetry
can be extended away from $p$ to 
a better approximation along a single 
null generator $\Gamma$ than across the
whole LCH. This calls to mind
null Fermi coordinates\cite{Blau:2006ar}, but we have used the
NNCs to describe the situation since they 
are better adapted to the LCH. 

Given that we confine the horizon to a narrow region surrounding the central horizon generator $\Gamma$, the integrals appearing in the Clausius relation will be dominated by their integrands evaluated on $\Gamma$. 
As such, any conditions that $\xi^a$ may need to satisfy in order for the Clausius relation to lead to a consistent equation of state will be conditions imposed on 
$\Gamma$. 

As 
motivated in Sec.~\ref{New choice}, 
the local Killing vector $\xi^a$ is taken to have a 
bifurcation surface $\Sigma_0$---or at least a bifurcation point---at $p_0$ 
to the past of $p$, where it vanishes and where its covariant derivative generates boosts in the plane orthogonal to the bifurcation surface. The NNCs of $p_0$ are $(U,V,x^A)=(0,V_0,0)$. It will be convenient to shift the affine parameter such that its origin coincides with the bifurcation point, i.e. we define $\tV \equiv  V - V_0$, so 
$p_0$ lies at $\tilde V = 0$.
We also want $\xi^a$ to be approximately tangent to the generator  $\Gamma$ that
connects $p_0$ to $p$,
so that the evolution of the LCH can be interpreted as a small perturbation
of a stationary background, which justifies the use of the Clausius
relation.\footnote{Moreover, 
if we 
were to 
let the local Killing vector and the tangent to the generator vary independently, it appears the Clausius relation 
would be 
overly restrictive.} 
 More specifically, the equation of state derivation will require that, at least to lowest order, $\xi^a$ coincides with $\tV k^a$, as would be the case on a Killing horizon.
And,  
finally, the derivation will require the Killing equation
and the Killing identity to hold at $O(\tV)$. 
Expressed in 
NNC components, the full set of requirements that must be imposed on $\xi^\a$ is:
\beq 
\left. \nabla_\a \xi_\b \right\vert_{p_0} 
&=& \left.(k_\a l_\b - l_\a k_\b)\right\vert_{p_0},
\label{boostcon}\\
\left. \xi^\a \right\vert_{\Gamma} &=& \tV \d_V^\a + O(\tV^2),
\label{xiGamma}\\
\left. \nabla_{(\a}\xi_{\b)} \right\vert_{\Gamma} &=& O(\tV^2),
\label{weak_Killingeqn}\\ 
\left. \nabla_\a \nabla_\b \xi_\c \right\vert_{\Gamma} &=&  
\left. \left(R^\d{}_{\a\b\c} \xi_\d \right)\right\vert_{\Gamma} +  O(\tV^2).
\label{Killing_identity}
\eeq

As
alluded to above, $\xi^a$ may be subjected to rather stronger conditions under which it approximates 
a true Killing vector 
more closely.  
In particular, we may choose $\xi^a$ such that the 
Killing identity (\ref{Kid}) holds {\it exactly} 
on $\Gamma$. This 
can be demonstrated 
by a perturbative argument 
which is given at the end of this section (see also \cite{RafNNC} for further details).
The identity then 
implies that the Killing equation also is exact on $\Gamma$,
provided it holds exactly at one point. 
Another consequence is that $\xi^a$ becomes exactly 
tangential to $\Gamma$, as will be shown next. Hence our claim is that all terms of quadratic or higher order in $\tV$ in (\ref{xiGamma}) - (\ref{Killing_identity}) may actually be set to zero.

If a vector field $\zeta^a$ satisfies the Killing identity along a 
geodesic with affine tangent vector $v^a$, it follows that
\beq \label{Killing consequence}
v^a \nabla_a \left( v^b \nabla_b \zeta_c \right) = R_{dabc} \zeta^d v^a v^b.
\eeq
Eqn.(\ref{Killing consequence}) is an ordinary differential equation of second order and has a unique solution along the geodesic, once $\zeta^a$ and $v^b \nabla_b \zeta^a$ are given at an initial point. It is easily verified that $s\,v^a$ is a solution to (\ref{Killing consequence}), 
where $s$ is any affine parameter on the geodesic, and $v^b \nabla_b(sv^a)=
\dot{s}v^a$. 
Hence if $\zeta^a=sv^a$ and $v^b \nabla_b\zeta^a=v^a$ at one point, 
it follows that $\zeta^a = s v^a$ everywhere on the geodesic. 
Now recall that $k^a$ is an affine tangent to $\Gamma$, and
the local Killing vector is chosen to vanish at $p_0$ and 
to satisfy $\left. \nabla_a \xi_b \right\vert_{p_0}=\left.(k_al_b - l_a k_b)\right\vert_{p_0}$. Taking $v^a=k^a$ 
thus yields $ \left. v^b \nabla_b \xi^a \right\vert_{p_0}= \left. v^a \right\vert_{p_0} $, so we may conclude that
$\xi^a=\tilde{V}k^a$ everywhere on $\Gamma$.\footnote{
It will generally be inconsistent with the Killing identity on $\Gamma$ to 
choose $\xi^a$ to be tangential to {\it all} the generators of a small horizon patch. 
Moreover, 
if the horizon expansion does not vanish at the bifurcation point,
such a choice would also be inconsistent with the Killing equation imposed at linear order in a spacetime neighborhood.}

As regards the  
Killing equation,
we recall that the NNC system is local inertial at $p$. As discussed at the beginning of the current section, it is then possible for the Killing equation (\ref{Keq}) to hold at $O(x)$, and this turns out to be consistent with the Killing identity on $\Gamma$. In NNCs, the approximate Killing equation implies that the components of the local Killing vector are of the form $\xi^\a = \tV \d_V^\a - U \d_U^\a + O(x^3)$.    

In summary, the main properties exhibited by our choice of local Killing vector are given by (\ref{boostcon}) together with
\beq
\left. \xi^\a \right\vert_{\Gamma} &=& \tV \d_V^\a ,
\label{strong_xiGamma}\\
 \nabla_{(\a}\xi_{\b)}  &=& O(x^2),
\label{Killingeqn}\\ 
\left. \nabla_\a \nabla_\b \xi_\c \right\vert_{\Gamma} &=&  
\left. \left(R^\d{}_{\a\b\c} \xi_\d \right)\right\vert_{\Gamma} .
\label{strong_Killing_identity}
\eeq

To conclude this section, we turn to the perturbative argument that the Killing identity (\ref{Kid}) can be satisfied exactly on $\Gamma$. We specify $\xi^a$ in a neighborhood of $\Gamma$ by its (covariant) components in NNCs as the Taylor series
\beq
\label{local_killing_NNC}
\xi_\a = U\d_\a^{V} - \tV \d_\a^{U}
+ C_{\b \c \a}\tx^\b\tx^\c 
+ D_{\b \c \d \a} \tx^\b\tx^\c\tx^\d + \dots
\eeq
where $\tx^\a=x^\a-V_0\d^\a_{V}$ and in particular $\tV=V-V_0$.
Similarly, on $\Gamma$ the deviation away from the Killing identity is written as a power series in the affine parameter $\tV$. As shown in \cite{RafNNC}, the latter series may be set to zero order by order through an appropriate choice of the expansion coefficients occurring in (\ref{local_killing_NNC}). 

More specifically, at the linear order (\ref{Killing_identity}) needed for the equation of state derivation, one finds
\begin{eqnarray}
C_{\a\b\c }&\sim& V_0\times \mbox{(Riemann components at $p$)},\\
D_{V \a\b \c } &\sim& \mbox{(Riemann components at $p$)}.
\end{eqnarray}
Thus the terms that are quadratic and cubic in  $\tilde{x}$ in the expansion (\ref{local_killing_NNC}) 
both contribute to $\xi_\a$ only at $O(x^3)$, counting $V_0$ as $O(x)$.
This is consistent with the Killing equation (\ref{Killingeqn}) at $O(x)$, from which one finds that quadratic order
terms must be absent in (\ref{local_killing_NNC}).

For the purpose of the general derivation of field equations all we need to know is
that the conditions (\ref{boostcon})-(\ref{Killing_identity}) can be met.
For the purpose of computing the actual entropy (which will have a dependence on the local Killing vector, see Sec. \ref{Noetheresque})
and comparing with the area in the GR case or with other expressions in higher derivative gravity, the detailed form of the local Killing vector is generally needed. For more explicit expressions of the higher order coefficients in (\ref{local_killing_NNC}) we refer to  \cite{RafNNC}.\\


\section{Black hole entropy as Noether charge} 
\label{Wald entropy}

In this section we review Wald's expression for 
black hole entropy in terms of the Noether 
charge\cite{Wald:1993nt,Iyer:1994ys}.
This expression will motivate the form of LCH entropy
that we adopt.

Consider a diffeomorphism-invariant Lagrangian field theory of gravity in arbitrary dimensions. A variation of the action is always equal to a sum of terms proportional to the field equations and a surface term. When the variation is induced by a vector field $\xi^a$, this equality can be expressed as the statement that a certain current $j^a$ is closed, i.e. $ \nabla_a j^a =0$. The current is closed even when evaluated off-shell and takes the form
\beq \label{closedcurrent}
j^a=\theta^a - L\, \xi^a - \left(2\,E^{ab}-T^{ab}\right)\xi_b + ...,
\eeq 
where the symplectic current $ \theta^a $ stems from the surface term, $L$ is the Lagrangian scalar in the action, $2\,E^{ab}-T^{ab}=0$ is the metric field equation and the dots indicate off-shell terms corresponding to the matter field equations. The Noether current $J^a$ is defined as $J^a=\theta^a - L \xi^a$, and on shell coincides with the closed current $j^a$. 
Furthermore, since the on-shell Noether current is closed
for any possible vector field $\xi^a$, 
it can be expressed       
as $J^a = 2 \nabla_b Q^{ab}$, where the antisymmetric tensor 
$ Q^{ab}$ is a local function of the fields and their derivatives\cite{Wald:1993}.
The tensor $Q^{ab}$ is
referred to as a Noether potential.\footnote{In this paper we 
use tensor notation. The formalism in Refs.~\cite{Wald:1993nt,Iyer:1994ys}
makes use of differential forms, which are dual to these
tensors. The relation between a $p$-form $F$
and the corresponding tensor $\tilde{F}$ is
$F_{a_1\dots a_p} = \tilde{F}^{b_1\dots b_{n-p}}\e_{b_1\dots b_{n-p}a_1\dots a_p}$, 
where $\e$ is the volume element.
The tensor corresponding to the exterior derivative 
$dF$ is $(n-p)\nabla_a\tilde{F}^{b_1\dots b_{n-p-1}a}$.}

Since the vector field $\xi^a$ only enters the expression for a 
Noether potential when taking Lie derivatives of tensor fields 
with respect to $\xi^a$, 
any Noether potential is of the form
\beq \label{Noetherform}
Q^{ab}[\xi] = W^{abc} \xi_c + P^{abcd} \nabla_{c} \xi_{d}, 
\eeq
apart from the possible addition of a total divergence.
More specifically,
\beq \label{potential}
Q^{ab}[\xi] = W^{abc} \xi_c + X^{abcd} \nabla_{[c} \xi_{d]} 
+ Y^{ab} + \nabla_c Z^{abc},
\eeq
where the tensors $W, X, Y, Z$ are locally constructed from the dynamical fields, $Y$ is linear in Lie derivatives with respect to $\xi$, and  $Z^{abc}$ is totally antisymmetric.
The decomposition of $Q$ in terms of $W, X, Y, Z$ is not
unique, and in addition the Noether potential has three sources of ambiguity,
coming from the freedom to add a total divergence to the Lagrangian, 
the symplectic potential, 
or the Noether potential itself. Using all this freedom
$Y$ and $Z$ can be set to zero, 
$X$ can be chosen~\cite{Iyer:1994ys} as
\beq \label{Noether}
X^{abcd}= - \frac{\partial L}{\partial R_{abcd}} + ...
\eeq
and $W$ is then given by~\cite{Lopes Cardoso:1998wt}
\beq\label{WMohaupt}
W^{abc}= 2 \nabla_d X^{abcd} + \mbox{matter terms} + ...,
\eeq
where
the dots indicate terms that stem from derivatives of the Riemann tensor 
in the Lagrangian.

If the theory admits stationary black hole solutions with a 
regular bifurcation surface,  
variations away from 
these solutions satisfy the first law (1), 
with the entropy $S_{BH}$ defined by \cite{Iyer:1994ys} 
\begin{equation} \label{generalform}
S_{BH} = \frac{2\pi}{\hbar}  \oint_{\Sigma}  Q^{ab}[\hat{\chi}] N_{ab} dA.
\end{equation}
Here, $\Sigma$ denotes any slice of the stationary horizon, 
$N_{ab}=2k_{[a}l_{b]}$ 
is its binormal (normalized as 
$N_{ab}N^{ab}=-2$),
$dA$ is the $(D-2)$ dimensional area element, and  
$\hat{\chi}^a$ is the horizon-generating Killing vector 
normalized to unit surface gravity. The integral is referred as the Noether charge and is invariant under the three sources of 
ambiguity~\cite{Jacobson:1993vj,Iyer:1994ys}. 
Invoking the stationary symmetry, it was further shown in these references that the entropy may equally be expressed by substituting $Q^{ab}[\hat{\chi}]$ in (\ref{generalform}) with  $X^{abcd} N_{cd}$. Thus the $W$ term does not contribute to the black hole entropy. 
For general relativity, $S_{BH}$ becomes the familiar 
Bekenstein-Hawking entropy $A/4\hbar G$. 


\section{Clausius relation and equation of state}
\label{Clausius LCH}

In this section we adopt the assumption that LCH's have an entropy of Noetheresque form, 
i.e. the entropy density
depends on the local Killing vector in the same way
that the Noether potential (\ref{Noetherform})
depends on the horizon generating Killing field.
With this entropy, we shall find that 
the Clausius relation applied to all LCHs, together with the local conservation
law for the matter stress tensor, can be satisfied provided that 
i) the entropy density can be identified with (the gravitational part of) a Noether potential of some Lagrangian, and
ii) the fields satisfy the metric field equation for that Lagrangian.\footnote{Note that for the equation of state to be well defined the matter stress tensor need not be the functional derivative with respect to the metric of an action.} This
is the equation of state for the corresponding entropy function.

\subsection{Clausius relation for a local causal horizon}
\label{Clausius}

This Clausius relation for a LCH is analogous to the first law for stationary black holes.
However, the latter relies on the stationarity of the black hole horizon,
as well as the fact that a horizon slice is a complete boundary component of a spacetime slice,
neither of which hold for a LCH. Nevertheless, it turns out that the 
construction can be localized enough to make the Clausius relation 
at least well defined.
First, as explained in Sec.~\ref{sec:slices} and Fig.~\ref{fig:slices},
we compare the entropy on two LCH slices $\Sigma$ and $\Sigma_0$
that share a common boundary,
so that together they form the boundary of a local patch $H$ of the horizon.
The lack of stationarity looks more problematic because,
for a dynamical black hole 
horizon in a general theory, there is no well 
defined notion of entropy available\footnote{General relativity
is an exception, where the area of the horizon slices is a good 
candidate for the dynamical entropy. This 
is motivated by the area theorem, as well as 
by the form of entanglement entropy; see 
Sec. \ref{Einstein equation of state}.}. 
The problem is that there is no Killing vector and, if a vector field is somehow selected,
the ambiguities in the definition of the Noether charge will make the entropy ambiguous,
unlike for a stationary black hole horizon.\footnote{Although the proposed dynamical entropy 
of Ref. \cite{Iyer:1994ys} (in terms of the instantaneous boost invariant metric) manages to bypass 
the ambiguities, it most likely does not satisfy a second law\cite{Jacobson:1995uq}.}
On the other hand, locally there is always an approximate Killing vector, 
the ``local Killing vector" constructed in Sec.~\ref{LKV}. This turns out to act enough like a
Killing vector to make the Clausius relation well defined in a certain local limit,
once a form for the entropy is adopted.

In the context of horizon thermodynamics it is natural to expect that
the entropy is an extensive quantity. Thus we assume that it can be expressed as an integral of a $(D - 2)$-form over a spacelike slice $\Sigma$ of the horizon. We adopt the dual description, and express the entropy as the integral
\beq \label{entropyform}
S =  \int_{\Sigma}  s^{ab} N_{ab} dA,
\eeq  
where the entropy density $s^{ab}$  is an antisymmetric tensor, $N^{ab}$ is the binormal to the slice and $dA$ is the area element on the slice.
The change of this entropy between the two horizon 
slices is
\beq\label{deltaS}
\delta S &=& S - S_0 \nonumber \\
&=&  \oint_{\Sigma \cup {\Sigma}_0}  s^{ab} N_{ab} dA \nonumber \\
&=&   2\int_{H} \nabla_b s^{ab} dH_a\nonumber\\
&=& -2\int_{H} \nabla_b s^{ab}\,  k_a dV dA,
\eeq
where $\Sigma_0$ is taken with the opposite orientation to $\Sigma$,
and in the third line we have used Stokes' theorem (for details
of Stokes' theorem on a null surface see Ref.~\cite{Poisson}). 

On the right hand side of the Clausius relation, $dS=\d Q/T$, 
we have the heat flux
\beq\label{dQ}
\d Q = \int_H (-T_{ab})\xi^b dH_a
\eeq
divided by the Unruh boost temperature $T=\hbar/2\pi$.
The heat flux integrand is proportional to the local Killing
vector which vanishes at $p_0$ and is thus of $O(x)$ in NNCs 
in the neighborhood of $p_0$.
The Clausius relation is imposed in the limit $p_0\rightarrow p$,
which means that the entropy change
integrand of (\ref{deltaS}), multiplied by $T$,  
must be equal to the $O(x)$ heat flux integrand, up to $O(x^2)$ terms. That is, 
\beq\label{ClausiusDivs}
- (\hbar/\pi)\nabla_b s^{ab}\,  k_a = T^{ab}\xi_b\, k_a + O(x^2).
\eeq
This relation is imposed at all points $p$ and for all null directions at $p$. 

We emphasize that the  limit $p_0\rightarrow p$
is taken in a formal sense only, in order to identify the
leading order contribution for a small region.
From a physical point of view, it makes no sense
to consider an arbitrarily small region, because 
quantum fluctuations of the metric presumably
invalidate our semiclassical considerations 
at sufficiently short distances.

\subsection{Entropy density of Noetheresque form}
\label{Noetheresque}

Inspired by
Refs. \cite{Parikh:2009qs, Padmanabhan:2009ry}, 
we propose a local entropy density 
of the same form as the Noether potential of Eqn. (\ref{Noetherform}), 
where $\xi^a$ is the local Killing vector, 
but a priori no restrictions are placed on 
$W$ and $P$, other than that they are constructed locally from the dynamical fields. 
We emphasize that, although we are using a similar notation, we do \textit{not} assume at this stage that the 
divergence of the local entropy density
is a Noether current associated to a Lagrangian, in contrast to the entropy density of stationary black holes.
This form of entropy density can be viewed as
the most general one that depends linearly on a local Killing vector that satisfies the Killing identity, and contributes
to the entropy change at $O(x^2)$. As mentioned in the
introduction, since the local Killing vector determines
the local notion of equilibrium and heat, it is not
entirely unnatural that the entropy would depend on
the Killing vector.  On the other hand, this entropy is
not strictly intrinsic to the horizon, and has no immediate
statistical interpretation that we are aware of.

We thus assume the entropy density takes the form
\beq \label{entropydensity}
s^{ab}=\frac{2\pi}{\hbar}Q^{ab},
\eeq
with
\beq \label{the proposal}   
Q^{ab}  &=& W^{abc} \xi_c + P^{abcd} \nabla_c \xi_d \nonumber \\
        &=& W^{abc} \xi_c + \left( X^{abcd}\, + \, Y^{abcd}\right) \nabla_c \xi_d.  
\eeq
In the second line of (\ref{the proposal}), the tensor $P$ is split into an antisymmetric part $X$ and a symmetric part $Y$:
\beq    
X^{abcd} = X^{ab[cd]} \quad \mathrm{and} \quad  Y^{abcd} = Y^{ab(cd)}.  
\eeq
We could of course absorb the factor of $2\pi/\hbar$
into the definition of $Q^{ab}$. The only reason we factor it
out here is so that the notation will make the 
analogy with the first law for stationary black hole horizons 
more transparent.

We now show that the symmetric part does not contribute to the Clausius relation.
The contribution of $Y^{abcd}$ to $\nabla_b Q^{ab}$ is 
\beq \label{sym part}
   \nabla_b  \left( Y^{abcd} \nabla_{(c} \xi_{d)} \right)= 
\nabla_b Y^{abcd}\, \nabla_{(c} \xi_{d)} + Y^{abcd} \nabla_b \nabla_c \xi_d.\nonumber\\
\eeq
Using the approximate local Killing equation (\ref{Killingeqn}),
we see that the first term on the RHS is of $O(x^2)$. 
The second term is symmetric in $cd$, so without
any change we may 
add a term in the Riemann tensor, yielding
\beq
 Y^{abcd} \left( \nabla_b \nabla_c \xi_d  - R^{f}_{\ph{f}bcd} \, \xi_f \right),
\eeq
which according to the approximate Killing identity 
(\ref{Killing_identity}) is of $O(x^A)$, where $x^A$ are the transverse
coordinates of the NNC system.
The contribution of (\ref{sym part}) is therefore
negligible in the limit of small, narrow horizon patches.  

Thus, at least for the part that contributes to the entropy
change, our proposal for the entropy density reduces to
$2\pi/\hbar$ times
\beq \label{lean proposal}   
Q^{ab} = W^{abc} \xi_c + X^{abcd} \nabla_c \xi_d, 
\eeq
where $W$ and $X$ are, so far, unspecified tensors, constructed locally from the dynamical fields, antisymmetric in the first two indices, and in addition $X$ is antisymmetric in the last two indices.  Note that the $X$ term is of $O(1)$, and the
$W$ term is of $O(x)$. Had we kept the $Y$ term it would have 
been of $O(x^2)$. 

Next we show that the Clausius relation requires that $W$ is a combination of divergences of $X$. The divergence of (\ref{lean proposal}) can be written as
\beq \label{divergenceofs}
\nabla_b Q^{ab}  & = & \left(\nabla_r W^{arb} + X^{arst} R^{b}_{\ph{b}rst} \right) \xi_b \nonumber \\
& + & X^{arst} \left( \nabla_r \nabla_s \xi_t  - R^{b}_{\ph{b}rst}\, \xi_b \right) \nonumber \\
& + & \left( W^{ast} + \nabla_r X^{arst} \right) \nabla_s \xi_t.
\eeq
Again, the first term is proportional to the local Killing vector
and hence of $O(x)$, while the second term involving the Killing identity may be neglected
on narrow patches.
 It is precisely here that the need to impose the Killing identity 
arises. Were this identity not satisfied by $\xi^a$, the second term
would make a contribution at the same order as the first term, which
would depend on the higher order coefficients
in the local Killing vector expansion. This would ruin our extraction
of the local field equation from the Clausius relation and would
probably make it impossible to consistently impose the Clausius relation
at all for such local Killing vectors.

In the third term, the symmetric 
part in $(st)$ is of $O(x^2)$ by virtue of the Killing equation, but a priori
the antisymmetric part is of $O(x^0)$. 
On the other hand, the integrand in the heat flux is of $O(x)$, so in order to be consistent with the Clausius relation (\ref{ClausiusDivs}), the $O(x^0)$ part of the 
third term must vanish. It follows that the local Clausius relation can only be 
satisfied at all spacetime points if $W$ and $X$ are related via
\beq\label{Wcondn}
W^{a[st]}+ \nabla_r X^{arst} = 0.    
\eeq
Because $W$ is antisymmetric in its first two indices, 
this relation 
completely determines $W$ as a function of $X$, to wit
\beq\label{W}
W^{arb} =  \nabla_s  \left( X^{sarb} + X^{sbra} +  X^{srba} \right).     
\eeq
Thus only the first term in (\ref{divergenceofs}) survives.

Notice that although the $W$ term does not contribute
to the entropy at $p_0$, nor at $p$ 
in the limit $p_0\rightarrow p$, its rate of change is
comparable to that of the $X$ term, so that---in contrast to the stationary comparison version of the first law of black hole mechanics~\cite{Iyer:1994ys}---it makes an
important contribution to the Clausius relation, which
could not be satisfied without 
it.\footnote{
One might think that perhaps a total divergence can be added 
to the entropy density such that the $W$ term is cancelled.
That is, if a tensor $Z^{abcd} = Z^{[abc]d}$ could be found such that
$W^{abd}=\nabla_c Z^{abcd}$, then we would have 
$Q^{ab} = (X^{abcd}-Z^{abcd})\nabla_c\xi_d + \nabla_c(Z^{abcd}\xi_d)$.
The total divergence term would not affect the changes in the entropy
from one slice to another with common boundary, so for the Clausius relation
this would be equivalent to having $W=0$ and replacing $X$ by $X-Z$.
The existence of such a $Z$ is at first glance
conceivable since, according to (\ref{W}), 
$W$ must be a divergence. However, the required total antisymmetry 
of $Z$ is not shared by the combination of $X$'s in (\ref{W}).
Only the divergences of $Z$ and $X$ enter the equation, but still it seems 
unlikely that such a $Z$ exists in general.}

\subsection{Equation of state}
\label{eos}
Substituting (\ref{W}) 
into 
the first term of
(\ref{divergenceofs}), and 
using the fact that
$\xi_a\propto k_a$ 
on the central generator, the validity of the 
Clausius relation (\ref{ClausiusDivs}) for all $k^a$ implies 
\beq \label{eom}
 R^{(a}_{\ph{a}rst} X^{b)rst}  + 2 \nabla_r \nabla_s X^{(a|s|b)r} + \Phi g^{ab} 
= -  \frac12 T^{ab} ,
\eeq
where $\Phi$ is some scalar function that may depend on the metric and curvature. 
Note that (\ref{eom})
follows from the Clausius relation irrespective of the values of expansion and shear anywhere on the horizon patch.

As was done in Sec. \ref{Impossibility}, we impose local conservation of energy-momentum to 
determine the function $\Phi$. Then Eqn.~(\ref{eom}) leads to
\beq\label{int}
\nabla^{a} \Phi  = -\nabla_b \left(R^{(a}_{\ph{a}rst} X^{b)rst}  
+ 2 \nabla_r \nabla_s X^{(a|s|b)r}  \right). \label{conditionL}
\eeq
In order for such a $\Phi$ to exist, the right hand side 
must be the gradient of a scalar. This integrability
condition further constrains the nature of $X$,
which so far is only required to be anti-symmetric in 
both the first and second pair of indices.

If the integrability condition is satisfied, then the 
left hand side of (\ref{eom}) is a divergence free tensor
constructed from the metric and its derivatives.
One way to obtain such a tensor is from the variational
derivative of a scalar action functional with respect to the
metric, $\d I_g[g]/\d g_{ab}$, 
which is automatically divergence free. In fact, it was argued
in Ref.~\cite{Curiel:2009as} 
that {\it all} such tensors arise in this way.
If so, then the `Clausius equation' (\ref{eom}) is precisely the equation of 
motion that derives from the action $I_g + I_{\rm matter}$,
with an undetermined cosmological constant.

\subsubsection{Relation between entropy density and Lagrangian}

We can be more specific about the relation between the
entropy and an action whose equation of motion is
(\ref{eom}). If a gravitational Lagrangian
$L[g_{ab},R^{a}_{\ph{a}bcd}]$ is a scalar formed 
algebraically from the metric
and Riemann tensor, then the corresponding equation of motion 
is precisely (\ref{eom}),
\footnote{To show this is a bit tricky.
One can regard the Lagrangian as
a function of $g_{ab}$ and $R_{abcd}$, and use the
identity $\partial L/\partial g_{ab} = -2R^{(a}_{\ph{a}rst}\d^{b)}_q \partial L/\partial R_{qrst}$.
This identity can be easily established by exploiting the fact that the Lagrangian can also be
considered as a function of the $(2,2)$ tensor $R^{ab}_{\ph{ab}st}$, without any
explicit dependence on the metric or $R_{qrst}$,
and taking the partial derivatives using the chain rule and $R^{ab}_{\ph{ab}st}=g^{aq}g^{br}R_{qrst}$. For a detailed derivation, see Ref.~\cite{Padmanabhan:2011ex}.}
with 
\beq \label{conditionin}
X^{abcd} = -\frac{\partial L}{\partial R_{abcd}},
\eeq
and $\Phi=L/2$.
Note that if $X$ is assumed to have this form, it has all the 
symmetries of the Riemann tensor. Using these symmetries 
one can show that our condition (\ref{W}) that determines $W$ in terms
of $X$ becomes exactly the same as
Eqn. (\ref{WMohaupt}).
Thus, with the sufficient condition 
(\ref{conditionin}) assumed, our conjectured entropy density is nothing but the specific choice discussed in Sec.~\ref{Wald entropy} 
for the Noether potential associated to a Lagrangian 
$L[g_{ab},R^{a}_{\ph{a}bcd}]$.
\footnote{We note that not all Noether potentials associated to a given Lagrangian satisfy the condition (\ref{W}), it 
is curious that the local Clausius relation restricts the form of the Noether potential in this manner. Perhaps this stems from the lack of exact symmetry in a general spacetime background.}

For a Noetheresque entropy density with $X$ of the form (\ref{conditionin}), 
and $W$ satisfying (\ref{Wcondn}), 
the divergence in 
Eq. (\ref{divergenceofs}) is nothing but the closed current (\ref{closedcurrent}) 
of Sec. \ref{Wald entropy}, for the Lagrangian scalar $L[g_{ab}, R^a_{\ph{a}bcd}]$. That is,  
\beq
2 \nabla_b Q^{ab} = \theta^a - L \xi^a - 2 E^{ab} \xi_b,
\eeq
where $E^{ab}$ is the variational derivative of $L$, 
given by the LHS of eqn (\ref{eom}). 
Specifically, the first term of the RHS of (\ref{divergenceofs}) is given by 
$- L \xi^a - 2 E^{ab} \xi_b$, while the symplectic current $\theta^a$ 
is given by the last two terms. 
We further note that the symplectic current vanishes for a Killing vector, while it was found here to be negligible for the local Killing vector. Similar remarks were also made 
in~\cite{Padmanabhan:2009ry}. 

Finally, we point out that this particular entropy density does not include a dependence on matter fields. However, such a dependence should perhaps not be ruled out a priori.

\subsection{Examples}

\subsubsection{$L(R)$ theories}

We now illustrate our result by a simple example, 
with $X$ given by
\beq\label{Xf}
X^{abcd} = \frac{f}{2} \left( g^{ac}g^{bd} - g^{ad} g^{bc}\right),
\eeq
where $f = f(R,R_{ij}, R_{ijkl})$ is an arbitrary scalar function. This expression satisfies all the symmetry requirements imposed so far, 
but from Eq. (\ref{conditionL}) we obtain
\beq
\nabla^{a} \Phi = - \frac{1}{2} f  \nabla^a R,
\eeq
which can not be integrated unless the function 
$f$ depends only on the Ricci scalar. 
Thus, we find a strong restriction on the form
of the entropy density if the Clausius relation
is to be consistent with energy conservation.
For example, the leading order term in the entropy
density can not be, say, $1+\a R_{ab}R^{ab}$.

If $f$ is only a function of $R$, it can
be written as $f=-dL/dR$ for some
function $L(R)$, and then we have $\Phi=L/2$
(the arbitrary additive constant freedom in 
$\Phi$ can be absorbed into $L$, and corresponds
to a cosmological constant).
With $f$ of this form, Eq.~(\ref{eom}) becomes
\beq
L'(R)  R_{ab} &-& \nabla_a \nabla_b L'(R) \nonumber \\ &+&\left(\Box L'(R) -\frac{1}{2} L\right) g_{ab} =\frac{1}{2}T_{ab}.
\eeq
This is identical to the equation of motion that results 
from the Lagrangian $L(R)+L_{\rm matter}$.  
In the case of GR, $L=R/(16\pi G)$, and we obtain the
usual Einstein equation. 

In the case of generic $L(R)$ theories,
we have thus obtained the field equation as an 
equation of state from the
Clausius relation, without the need for a term representing
internal entropy production due to bulk viscosity, unlike in 
the non-equilibrium framework of 
Ref.~\cite{Eling:2006aw}. As described in Sec.~\ref{Impossibility}, in that approach the 
change of the entropy density 
$f(R)$ coming from the gradient of the
background $R$ at the equilibrium point is 
cancelled by a choice of non-zero horizon 
expansion, which then leads to the bulk viscosity 
term in the entropy balance equation. 
Instead, with the Noether charge entropy
density, the $W$ term is chosen so that 
the Clausius relation can be satisfied without
internal entropy production. 

\subsubsection{Lovelock theories}

Suppose that $X^{abcd}$ has no higher than second 
derivatives, and that 
also $\nabla_a X^{abcd} =0$, so (\ref{W})
implies $W^{abc}=0$. Then it follows that  
the tensor on the left hand side of the 
field equation ({\ref{eom}) is second order in derivatives.
Moreover, after imposing the integrability condition
(\ref{int}), we infer that this tensor, built from the metric and its
first and second derivatives, must be identically divergence free.
The only such tensors come from the metric variation of a 
Lovelock Lagrangian\cite{Lovelock}, so the assumed properties 
of such an $X^{abcd}$, together with the Clausius relation,
imply that $X^{abcd}$ arises as in 
(\ref{conditionin}) from a Lovelock 
Lagrangian of the form $L =L[g_{ab},R^a_{\ph{a}bcd}]$.

\subsubsection{General relativity}

In example (\ref{Xf}) with the choice $f = - (16 \pi G)^{-1}$, Eqn. (\ref{W}) implies that $W=0$. The corresponding entropy density obtains from Eqns. (\ref{lean proposal}) and (\ref{entropydensity}) and reads $s_{GR}^{ab} = - (8 \hbar G)^{-1} \nabla^{[a} \xi^{b]}$. The steps of Sec. \ref{eos} then lead to the Einstein equation of state.

For the Killing vector (\ref{local_killing_NNC}) we have $ \nabla^{[a} \xi^{b]}= 2 k^{[a}l^{b]}+ O(x^2)$.
Any horizon slice has a local binormal of the form $N_{ab}=2 k_{[a}l_{b]} +  k_{[a}m_{b]}$ where
$m^b$ is some spacelike vector tangent to the horizon and therefore orthogonal to $k^a$. 
For generic slices, the resulting entropy (\ref{entropyform}) 
coincides to leading order with the area in units of $1/4\hbar G$.

The \textit{difference} in entropy between two slices requires a more careful analysis. 
On a slice of constant affine parameter $V$, the entropy will differ from the area at subleading order, as it does on any slice. The coefficients of the terms of subleading order are given in terms of the coefficients $C$ and $D$ of Eqn. (\ref{local_killing_NNC}), many of which are in turn set by the Killing identity (\ref{Killing_identity}) imposed on $\Gamma$. These coefficients can be shown to be such that the subleading order terms, in the case of a slice of constant $V$, involve only transverse coordinates. Therefore, when comparing two slices of constant affine parameter, the subleading terms drop out and the entropy difference to leading order coincides with the area difference.  

However, for the Clausius relation of Sec. \ref{Clausius}, we need to compare two slices with common boundary.
In that case, on at least one of these slices the mismatch between entropy and area at subleading order does depend on $V$, in such a way that the entropy and area differences will not generically coincide to leading order,
 unlike in the first law of global horizon mechanics in GR
or the original Einstein equation of state derivation~\cite{Jacobson:1995ab}.  
The easiest way to see this is to employ Stokes' theorem (\ref{deltaS}) with the entropy density 
$s_{GR}^{ab} = - (8 \hbar G)^{-1} \nabla^{[a} \xi^{b]}$. To evaluate the integrand of (\ref{deltaS}), we use (\ref{divergenceofs}) and recall that the approximate Killing vector was chosen such that (\ref{divergenceofs}) is dominated by its first term. In this way we can express the entropy difference as
\beq\label{GR_entropy_diff}
\delta S \approx  \int_{H}  R_{ab}\,  k^a k^b \, \left(V-V_0\right) dV dA,
\eeq
where the integration range $H$ is a patch of the LCH enclosed by two slices. 
This integral will be equal to the area difference to leading order
only for a very special choice of slices, as we now describe.

Suppose that the first slice $\Sigma_0$ is the surface $V=V_0$, 
and the shape of the second slice $\Sigma$ is such 
that the enclosed horizon patch nearly amounts to the 
entire interval $\left[V_0,0 \right]$ (see Fig.~\ref{GR_Slice}). 
\begin{figure}
\centering
\includegraphics[width=8cm]{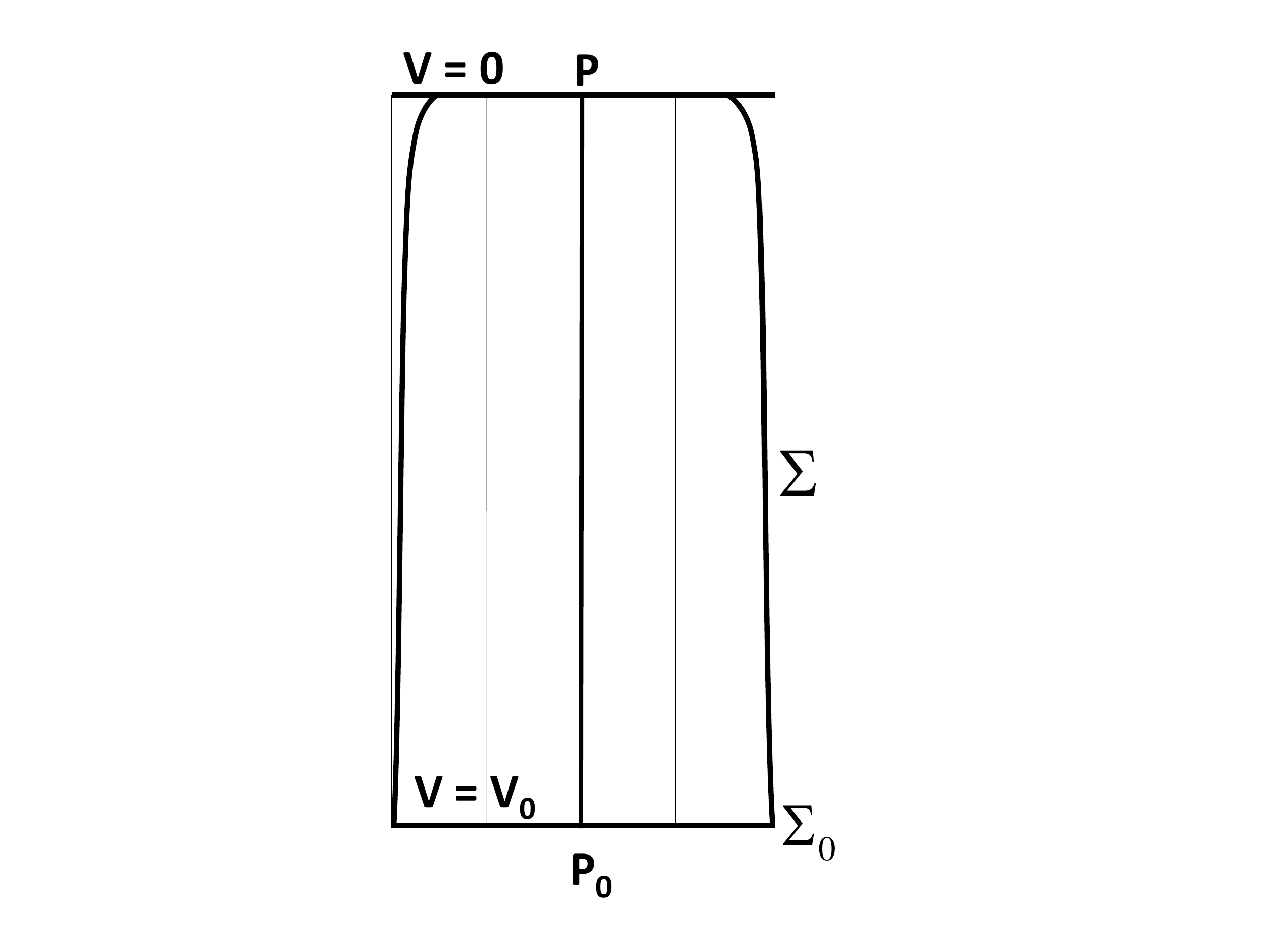}
\caption{\label{GR_Slice}
 A local causal horizon patch enclosed by 
a special pair of slices.
The entropy change
associated with $s_{GR}^{ab} = - (8 \hbar G)^{-1} \nabla^{[a}\xi^{b]}$ 
is equal at leading order to the area change over $4 \hbar G$ 
if ({\it i}) the horizon expansion and shear vanish at $p$, and 
({\it ii}) the patch between the slices $\Sigma_0$ and 
$\Sigma$ nearly coincides with the interval $[V_0,0]$ across the
entire patch.}
\end{figure}
Then, as indicated in (\ref{generator_integral}), the entropy difference may equivalently be approximated by $\delta S \approx  \int_{H}  R_{ab}  k^a k^b \, V dV dA$. If in addition the expansion and shear of the horizon generators vanish at $p$, 
then the Raychaudhuri equation tells us that 
$\delta S \approx  \int_{H}  \theta\, dV dA$.
Since the integration range is nearly the entire interval $\left[V_0,0 \right]$
this evaluates to $\delta S \approx A(V=0)-A(V=V_0)$, where 
$A(V)$ denotes the area of a slice of constant affine parameter $V$. The area of 
$\Sigma$ nearly coincides with $A(V=0)$ because its steep sides are ``nearly null"
and hence contribute negligibly to the area. Clearly, if the two slices are not chosen in 
this special fashion the entropy difference will not coincide to leading order with the area difference.

Nevertheless, if the entropy difference between two generic slices with common boundary obeys the local Clausius relation, the Einstein equation must follow as a consequence.  
We can illustrate the way this works with a trivial example in flat spacetime. A small patch of a light cone may be considered to be part of a LCH, enclosed by two spacelike slices with common boundary. The boost energy flux is identically zero, so the Clausius relation tells us the entropy on both slices must be identical. Employing $s_{GR}^{ab}$, this may be easily verified. However, the area of these slices is clearly not identical.
  
The entropy density $s_{GR}^{ab}$ also shows it is not only thermodynamically more natural to place the bifurcation point of the local Killing vector field
to the past of the terminal point (rather than coincident with the terminal point) as was discussed in Sec.~\ref{New choice}, it is actually required if we want the contribution to the entropy on a constant $V$ slice in the GR case to correspond to the area rather than minus the area.

To see why this is so, consider the fact that in the local Noether
charge approach we have derived the Einstein equation as an
equation of state, without specifying where the bifurcation point
lies. However, if we want the entropy to agree with the area
(to leading order), then $s^{ab}N_{ab}$ must be a positive constant.
In the case of GR, the entropy density is
$s_{ab}\propto -\nab_{[a}\xi_{b]}\sim - \nab_a\xi_b$, so
$s^{ab}N_{ab}\sim - k^a l^b \nab_a \xi_b$.
Since to leading order the causal horizon is a Killing horizon,
we have $k^a \nab_a \xi_b\propto  \pm \xi_b\propto \pm k_b$,
where the $+$ sign holds when the bifurcation point lies to the
past, so that the Killing vector is stretching in the future direction,
and the $-$ sign holds when the bifurcation point lies to the future,
so the Killing vector is shrinking (recall that we take the Killing vector
to be future pointing on the horizon).
Hence  $s^{ab}N_{ab}\sim \mp  l^b k_b =\pm 1$. To have a positive
entropy we must therefore take the bifurcation
point in the past.  We have no option to reverse the sign of the
assumed entropy density, since that would lead to a field equation
with the opposite sign for the gravitational constant. That is,
a negative entropy change would arise from a positive heat flux.

On the other hand, choosing the local Killing vector in this way entails a disturbing feature. If the proposed entropy represents the state of the horizon, one would expect it, and its difference,  to depend only on quantities intrinsic to the horizon. However, the entropy difference is given by the boost energy flux, which depends on the local Killing vector, which in turn depends on a freely chosen location $V_0$ of the bifurcation surface $\Sigma_0$. 
This is unsatisfying, as $V_0$ is a freely chosen parameter and does not encode any intrinsic property of the LCH.

Finally, we  point out that the entropy density $s_{GR}^{ab}$ is only 
one among 
an infinite number 
to give rise to the Einstein equation since, by Stokes' theorem, addition to the entropy density of any total divergence will have no influence on the entropy difference
This is in keeping with the fact that the Clausius relation is a thermodynamic relation, from which the value of the statistical entropy can only be deduced up to addition of an arbitrary constant.

\subsection{Including derivatives of curvature}
\label{nabla R} 
On general grounds, one may expect strong gravitational fields to be governed by a field equation derived  from a Lagrangian that includes covariant derivatives of the Riemann tensor. In the present context, this raises the question whether such a field equation may arise as an equation of state corresponding to an entropy density of the form (\ref{lean proposal}), introduced in Sec. {\ref{Noetheresque}}. As was discussed in that section, in order for this to be possible the tensors $W$ and $X$ must be related by Eqn. (\ref{W}).   This may be possible, but we point out that the specific choice of Sec. \ref{Wald entropy} for the Noether potential of such a Lagrangian does not satisfy the relation (\ref{W}). The problem is that the divergence of this Noether potential contains a Lie derivative of the Riemann tensor with respect to the local Killing vector. Whereas such a derivative would vanish for a true Killing vector, it must be expected to be of order one in a general  spacetime. Furthermore, it does not appear that relation (\ref{W})
can be salvaged by appealing to the freedom to add total divergences at the different stages in the construction of the Noether potential. If this is true it would mean that the strict analogy to the first law of black hole mechanics breaks down, since that law applies for any diffeomorphism-invariant Lagrangian. 

However, it would still leave open the possibility that entropy densities of the form (\ref{lean proposal}) exist that do satisfy relation (\ref{W}) and  give rise to the field equations of high derivative Lagrangians, but that are not Noether potentials associated with such Lagrangians. As shown in the Appendix,
the integrability condition that must be satisfied in order for such entropy densities to exist (whether or not they are Noether potentials) may be written in the form of a pair of tensor equations of first order, at least for Lagrangians that depend on no more than first order covariant derivatives of the Riemann tensor.

\section {Conclusion}

The question we addressed is whether gravitational field equations with higher
derivative terms can be derived from the Clausius relation applied to a higher
derivative horizon entropy. First we discussed problems that arise with previous
approaches to this problem. Then we adapted the starting point of one of those approaches, and assumed
that horizon entropy depends on an approximate local Killing vector in a way that
mimics
the diffeomorphism Noether charge
that yields the entropy of a stationary black hole.
We showed that the problems can all be avoided by a careful choice of the nature of the horizon patch to which the
Clausius relation is applied. In particular, the Clausius relation must refer to the change of
entropy between two slices of the horizon that together form the complete boundary of a
patch, and this patch must be narrow enough to neglect violations of the Killing identity.
We exploited a power series expansion in a coordinate
system adapted to the horizon to establish the required properties of the
local Killing vector.

Together with matter energy conservation, the Clausius relation applied to all such
local horizon patches leads to an integrability condition on the assumed horizon entropy
density. We showed that this condition can be satisfied if the latter is in fact a
Noether potential 
associated with
a Lagrangian 
constructed algebraically from
the metric and Riemann tensor.  
In that case the Clausius relation implies that the
field equation for that Lagrangian holds. We have not proved that this is the only way
to satisfy the integrability condition, but that may be the case. In particular, the field equation for a theory
with derivatives of curvature in the Lagrangian is unlikely to be obtained in this way using for the entropy
density a Noether potential derived from the Lagrangian, although it might conceivably arise from a different 
entropy density.

The higher derivative extension of the equation of state derivation was 
achieved in this paper at the 
cost of 
introducing a dependence of the entropy on
the choice of local Killing vector. Whereas this dependence occurs at subleading order, the entropy difference depends on the local Killing vector at leading order.
It is therefore not clear to us whether
the derivation has any thermodynamic significance. We regard it as a 
positive answer to a technical 
question,
but its physical interpretation remains obscure. 
Perhaps 
the steps we have taken are a valid part of a
picture that will only become clear once so far overlooked subtleties are taken into
account. For example, the arbitrariness of an additive constant in thermodynamic entropy,
the contribution from the entanglement entropy, and the related need to regularize
by some kind of subtraction or comparison might play an important role
in formulating the local thermodynamics of the vacuum. Then again, it may
just be that the contribution of higher curvature terms to a gravitational field equation
cannot be sensibly captured at a local thermodynamic level.

\begin{acknowledgments}
We are grateful to R.~Brustein, M.~Hadad, T.~Padmanabhan and A.~C.~Wall
for helpful discussions. 
This work was supported in part by the National Science Foundation under Grant No. 
PHY-0903572.
\end{acknowledgments}

\appendix
\section{Integrability condition for equations of state corresponding to Lagrangians
$L[g_{ab}, R^a_{\,\,bcd}, \nabla_f R^a_{\,\,bcd}]$.}

For notational convenience, we define the tensor $Z$~\cite{Jacobson:1993vj} by
\beq
Z^{f:abcd} \equiv \frac{\partial L}{\partial \left(\nabla_f R_{abcd}\right)}.
\eeq
Then the choice of Noether potential of Sec. \ref{Wald entropy} for a Lagrangian $L[g_{ab}, R^a_{\ph{a}bcd}, \nabla_f R^a_{\ph{a}bcd}]$ is given by
\beq \label{canonicalchoice}   
Q_{(0)}^{ab} = W_{(0)}^{abc} \xi_c + X_{(0)}^{abcd} \nabla_c \xi_d, 
\eeq
where
\beq   
X_{(0)}^{abcd}= -\frac{\partial L}{\partial R_{abcd}} + \nabla_f Z^{f:abcd}, 
\eeq
\beq   
W_{(0)}^{abc}= 2 X_{(0)}^{abcd} + a:[bc] - b:[ac] - c:[ab]
\eeq
and
\beq
a:[bc] \equiv Z^{a:[b}_{\ph{a:[b}def}R^{c]def}.
\eeq
Due to the terms $a:[bc]$ etc., the relation (\ref{W}) is not satisfied for $Q_{(0)}^{ab}$. 
We now suppose that the entropy density consists of the Noether potential $Q_{(0)}^{ab}$ with a further Noetheresque potential added. That is, we take 
\beq \label{good}  
s^{ab}=2 \pi/\hbar \, Q^{ab},
\eeq
where $ Q^{ab}= Q_{(0)}^{ab}+Q_{(1)}^{ab}$ and $Q_{(1)}^{ab} = W_{(1)}^{abc} \xi_c + X_{(1)}^{abcd} \nabla_c \xi_d$.

The divergence of $Q_{(0)}^{ab}$ is given by
\beq \label{divQ0}
2\nabla_b Q_{(0)}^{ab} = \theta^a - L \xi^a -2 E^{ab} \xi_b,   
\eeq
where $E^{ab}$ is the variational derivative of $L$, the symplectic current reads
\beq
\theta^a 
& = &  Z^{a:bcde} \mathcal{L}_{\xi} R_{bcde} \nonumber \\ 
& + & 2 X_{(0)}^{abcd} \left( \nabla_b \nabla_c \xi_d  - R^{f}_{\ph{f}bcd} \, \xi_f \right) \nonumber \\
& + & \left( 4 \nabla_d X_{(0)}^{abcd}+ A^{abc}\right) \nabla_{(b} \xi_{c)}, 
\eeq
and $A^{abc}$ is a combination of terms of the form $ZR$. 
If the entropy density (\ref{good}) is to give rise to the field equation derived from $L$, the divergence of $Q_{(1)}^{ab}$ must take the form
\beq \label{divQ1}
2\nabla_b Q_{(1)}^{ab} & = & - Z^{a:bcde} \mathcal{L}_{\xi} R_{bcde} 
+ F \xi^a \nonumber \\
& + & F^{abcd}  \left( \nabla_b \nabla_c \xi_d  - R^{f}_{\ph{f}bcd} \, \xi_f \right)
\nonumber \\
& + & F^{abc} \nabla_{(b} \xi_{c)},
\eeq
for some tensors $F$. From the symmetries of the Riemann tensor and the definition of the Lie derivative, this can be seen to be equivalent to the system
\beq \label{system}
W_{(1)}^{a[bc]} + \nabla_d X_{(1)}^{adbc} = -4 Z^{a:[b}_{\ph{a:[b}def}R^{c]def}, \nonumber \\
\nabla_b W_{(1)}^{abc} + X_{(1)}^{abde}R^{c}_{\ph{c}bde} = 
- Z^{a:bdef}\,\nabla^c R^{b}_{\ph{b}def} + F g^{ac}.
\eeq
In conclusion, if entropy densities of the desired form exist, there must exist an integrating function $F$ such that (\ref{system}) has a solution. There is a relation between the integrating function $F$ and the integrating function $\Phi$ of Eqn. (\ref{eom}).
Namely, comparing (\ref{divQ0})-(\ref{divQ1}) with (\ref{divergenceofs}) reveals that in this case $\Phi= L/2 + F/2$.

\end{document}